\documentclass[
 reprint,
superscriptaddress,
nofootinbib,
 amsmath,amssymb,
 aps,
prd,
floatfix,
longbibliography
]{revtex4-2}

\usepackage{orcidlink}
\usepackage[english]{babel}
\usepackage{xcolor}
\usepackage{graphicx}
\usepackage{dcolumn}
\usepackage{bm}
\usepackage[separate-uncertainty=true,multi-part-units=single]{siunitx}
\usepackage{hyperref}
\hypersetup{
     colorlinks=true,
     breaklinks=true,
     linkcolor=blue,
     filecolor=blue,
     citecolor=blue,      
     urlcolor=cyan,
     }
\usepackage[caption=false]{subfig}

\newcommand{\Neff}{N_{\text{eff}} }
\newcommand{\maxyr}{max(\rho_Y/\rho_r) }
\newcommand{\YHe}{Y_{\text{He}} }
\newcommand{\etaBBN}{\eta_{\mathrm{BBN}} }
\DeclareSIUnit[quantity-product = {}]\parsec{\text{pc}}
\graphicspath{{./}{figures/}}

\begin{document}

\title{Was Entropy Conserved between BBN and Recombination?}

\author{Alexander C. Sobotka\,\orcidlink{0000-0002-7576-5417}}
\email{asobotka@live.unc.edu}
\affiliation{Department of Physics and Astronomy, University of North Carolina at Chapel Hill,\\
Phillips Hall CB3255, Chapel Hill, North Carolina 27599, USA }

\author{Adrienne L. Erickcek\,\orcidlink{0000-0002-0901-3591}}
\email{erickcek@physics.unc.edu}
\affiliation{Department of Physics and Astronomy, University of North Carolina at Chapel Hill,\\
Phillips Hall CB3255, Chapel Hill, North Carolina 27599, USA }

\author{Tristan L. Smith\,\orcidlink{0000-0003-2685-5405}}
\email{tsmith2@swarthmore.edu}
\affiliation{Department of Physics and Astronomy, Swarthmore College, Swarthmore,\\ Pennsylvania 19081, USA}

\date{\today}

\begin{abstract}
We test the assumption of entropy conservation between Big Bang nucleosynthesis and recombination by considering a massive particle that decays into a mixture of photons and other relativistic species. We employ \textit{Planck} temperature and polarization anisotropies, COBE/FIRAS spectral distortion bounds, and the observed primordial deuterium abundance to constrain these decay scenarios. If between $56\%$ and $71\%$ of the decaying particle's energy is transferred to photons, then $\Neff$ at recombination is minimally altered, and \textit{Planck} data alone allows for significant entropy injection. If photons are injected by the decay, the addition of spectral distortion bounds restricts the decay rate of the particle to be $\Gamma_Y > \SI{1.91e-6}{\per\second}$ at $95\%$ confidence level (C.L.). We find that constraints on the energy density of the decaying particle are significantly enhanced by the inclusion of bounds on the primordial deuterium abundance, allowing the particle to contribute at most $2.35\%$ ($95\%$ C.L.) of the energy density of the Universe before decaying.  
\end{abstract}
\maketitle

\section{Introduction} \label{sec:intro}
An underlying assumption of the standard cosmological model is that the comoving entropy density of relativistic species is conserved. However, many alternative scenarios include entropy injection into the thermal bath of relativistic particles. Extreme entropy injection is needed to completely repopulate the relativistic bath after an early matter-dominated era \cite{pantano_dynamics_1993,chung_superheavy_1998,giudice_largest_2000,allahverdi_production_2002}. Minor entropy injections that only make small alterations to the radiation density have been proposed to alter the relic dark matter abundance after thermal freeze-out \cite{asaka_opening_2006,patwardhan_diluted_2015}, change the pre-recombination expansion history \cite{ichikawa_increasing_2007,fischler_dark_2011,hasenkamp_dark_2012}, and even relax the Hubble tension \cite{aboubrahim_analyzing_2022, aloni_step_2022}, which is a discrepancy between the present-day expansion rate inferred from local measurements and that inferred from the cosmic microwave background (CMB) \citep{bernal_trouble_2016,schoneberg_h_0_2021,di_valentino_realm_2021}.

Significant entropy injections, such as those used to transition out of an early matter-dominated era, must be complete before neutrino decoupling in order to avoid altering the light-element abundances predicted by standard Big Bang nucleosynthesis (BBN) and impacting the CMB anisotropies. Therefore, significant entropy injection is generally constrained to have completed at temperatures hotter than about 4 MeV \cite{de_salas_bounds_2015,hasegawa_mev-scale_2019}.

Minor entropy injections can occur during or after BBN, but are heavily constrained by their influence on the expansion rate, baryon-to-photon ratio, and photodisintegration of light nuclei  \cite{lindley_cosmological_1985,scherrer_primordial_1988,scherrer_primordial_1988-1,frieman_eternal_1990,protheroe_cascade_1995,hufnagel_bbn_2018,kawasaki_big-bang_2020}. Such scenarios have been thoroughly considered in the context of decaying axion-like particles \cite{cadamuro_cosmological_2012,millea_new_2015,depta_robust_2020,balazs_cosmological_2022} for which observations of primordial light-element abundances and the CMB provide stringent constraints. However, these investigations focus on specific axion models and therefore cannot place comprehensive constraints on general entropy injection.

In this work, we test the assumption of entropy conservation between BBN and recombination by considering a generic massive particle that decays into a mixture of photons and other relativistic species (e.g.\ dark radiation). We explore the bounds that can be placed on the particle decay rate, $\Gamma_Y$, and the contribution that the particle makes to the energy composition of the Universe before it decays. Since these constraints depend on what relativistic species this massive particle decays into, we also explore what bounds can be placed on $f_\gamma$, the fraction of the decaying particle's energy that is transferred to photons. 

We restrict our analysis to particles with rest-energies less than $\SI{3.2}{MeV}$ such that the maximum energy of any decay products is less than the binding energy of beryllium \mbox{($\SI{1.59}{MeV}$)}. Photons generated by decaying particles with larger masses would photodisintegrate light nuclei and alter the abundances of primordial elements such as deuterium, helium, and lithium. Photodisintegration of deuterium has been shown to place very stringent constraints on such entropy injections \cite{hufnagel_bbn_2018,kawasaki_big-bang_2020,poulin_non-universal_2015,kawasaki_revisiting_2018,forestell_limits_2019,poulin_cosmological_2017,balazs_cosmological_2022}. We note that the injection of photons with energies between $1.59$ and $\SI{2.22}{MeV}$ would destroy beryllium and not deuterium, leading to a potential solution to the lithium problem (e.g.\ \cite{kusakabe_big-bang_2013}). Although the inference of the primordial lithium abundance from observations of metal-poor stars is subject to uncertainties in stellar modeling \citep{korn_probable_2006,sbordone_metal-poor_2010}, these observations do establish a minimum primordial lithium abundance, and therefore rule out energy injections that cause the lithium abundance to fall below this level. We do not need to consider such bounds under the assumption that the mass of the decaying particle is less than $\SI{3.2}{MeV}$.


For a particle that is in kinetic equilibrium with the Standard Model to not contribute to the photodisintegration of beryllium \textit{and} be non-relativistic
at temperatures less than $\SI{1}{MeV}$, it must have a mass in the range of \mbox{$\SI{2.7}{MeV} \lesssim m \lesssim \SI{3.2}{MeV}$}.\footnote{A massive boson transitions from evolving as $a^{-4}$ to $a^{-3}$ at a pivot temperature of $T_{p} = m/2.7$ \citep{ganjoo_effects_2022}.} However, this tight mass window can be relaxed if the decaying particle is part of a hidden sector that is not thermally coupled to the Standard Model \cite{kolb_shadow_1985,berezhiani_asymmetric_1996,feng_thermal_2008,adshead_chilly_2016,adshead_reheating_2019}. Hidden sector particles could decay into dark radiation \cite{anchordoqui_decaying_2021, nygaard_updated_2021} or Standard Model particles \cite{zhang_long-lived_2015,berlin_thermal_2016,dror_co-decaying_2016}. 

If the hidden sector was originally in kinetic equilibrium with the Standard Model particles and then decoupled while all of the Standard Model particles were relativistic, the hidden sector could be significantly colder than the visible sector. For example, \mbox{$T_{\mathrm{HS}}/T_{\mathrm{SM}} \approx 0.465$} when \mbox{$T_{\mathrm{SM}} = \SI{1}{MeV}$} if the $Y$ particles are the only relativistic component of the hidden sector at decoupling. In this case, a hidden sector boson with a mass of \mbox{$\SI{1.3}{MeV} \lesssim m < \SI{3.2}{MeV}$} would be non-relativistic at proton-neutron freeze-out \textit{and} be light enough such that any entropy injection from the decay would not result in photodisintegration of beryllium. It is also possible that the hidden sector was never in kinetic equilibrium with the Standard Model, in which case the hidden sector could be arbitrarily cold. We assume the hidden sector to be sufficiently cold such that the decaying particle is non-relativistic by neutron-proton freeze-out. Under this assumption, our calculations do not depend on the specific mass of the decaying particle.

To constrain a generic decaying particle we employ \textit{Planck} measurements of the CMB temperature and polarization anisotropies \cite{planck_collaboration_planck_2020}, the most recent measurement of the primordial deuterium abundance \cite{cooke_one_2018}, and the COBE/FIRAS limit on spectral distortions \cite{mather_measurement_1994,fixsen_cosmic_1996}. Entropy injection before recombination will alter the effective number of relativistic species, $\Neff$, resulting in a change in the expansion rate. Alterations to the pre-recombination expansion history may be constrained by observations of CMB anisotropies \cite{samsing_model_2012}. Specifically, altering the expansion rate via $\Neff$ affects the amplitude of small scale perturbations \cite{hou_how_2013,planck_collaboration_planck_2014} and introduces a phase shift to the baryon-photon acoustic oscillations \citep{bashinsky_signatures_2004, follin_first_2015,baumann_phases_2016}. Observations of the primordial deuterium abundance are an excellent way to provide additional constraining power on nonstandard models, with measurements now reaching 1\% precision \cite{cooke_one_2018}. Finally, scenarios that inject photons before recombination may be constrained by CMB spectral distortions \cite{chluba_teasing_2014, lucca_synergy_2020, bolliet_spectral_2021}. 

We combine the constraining power of CMB anisotropy data, deuterium measurements, and CMB spectral distortions by employing a Markov Chain Monte Carlo (MCMC) analysis. In doing so, we derive bounds on the amount and type of radiation that can be injected by a decaying hidden sector particle. 

In many ways, our investigation generalizes the work of \textcite{millea_new_2015} who considered 
non-photodisintegrating injections from axion-like particles that decay solely into photons between BBN and recombination. The specific coupling of these axion-like particles to photons restricts the possible parameter space that can be explored. We remain agnostic regarding a specific particle model. In doing so, this work has three key distinctions from \textcite{millea_new_2015}: (1) we are able to derive a broadly applicable bound on the maximum level of entropy injection allowed between BBN and recombination, (2) we can consider a particle that decays into photons \textit{and} dark radiation, and (3) our full analysis of CMB anisotropies allows us to investigate the effects that entropy injection has on standard cosmological parameters, including $H_0$. We determine that minor entropy injection from a hidden sector particle is not a promising avenue for addressing the Hubble tension. A more exotic change to the Universe's radiation content, such as the inclusion of strongly interacting radiation \citep{aloni_step_2022}, is needed to significantly increase the value of $H_0$ inferred from observations of the CMB.

The structure of this paper is as follows. In Sec.\ \ref{sec:model}, we describe the model for our decaying particle. In Sec.\ \ref{sec:Effects}, we explore the primary effects that the decay has on $\Neff$ and the primordial abundances of helium and deuterium, and we review the expected constraints from spectral distortions. We outline the choice of priors and likelihood functions for our MCMC analysis in Sec.\ \ref{sec:MCMC}, and Sec.\ \ref{sec:results} discusses the results of these analyses. We conclude with a summary in Sec.\ \ref{sec:conclusions}, and we include an Appendix that contains a derivation of our analytical model and technical details of our modifications to the public code known as CLASS (Appendix \ref{sec:CLASSmodifications}), a derivation for the post-decay $\Neff$ (Appendix \ref{sec:postdecay}), and supplemental MCMC results (Appendix \ref{sec:MCMCruns}). 

\section{Decaying Particle Model} \label{sec:model}

We consider a sub-dominant hidden sector $Y$ particle that decays into photons and other relativistic particles (e.g.\ dark radiation) sometime between BBN and recombination. The Eqs.\ governing the background evolution for the energy density of the $Y$ particle ($\rho_Y$), photons ($\rho_\gamma$), and other ultra-relativistic species ($\rho_{ur}$), are
\begin{align}
   \frac{d}{dt} \rho_Y +3H\rho_Y &= -\Gamma_Y \rho_Y  \label{eq:rho_Y}\,, \\
    \frac{d}{dt} \rho_\gamma + 4H\rho_\gamma &= f_\gamma\Gamma_Y \rho_Y \,,\label{eq:rho_g} \\
    \frac{d}{dt} \rho_{ur} + 4H\rho_{ur} &= (1-f_\gamma)\Gamma_Y \rho_Y \,, \label{eq:rho_nu}
\end{align}
where $\Gamma_Y$ is the decay rate of the $Y$ particle, $H\equiv \dot{a}/a$ is the Hubble parameter (with the over-dot referring to a proper time derivative), and $f_\gamma$ dictates what fraction of the decaying particle's energy is transferred to photons.\footnote{More precisely, the terms on the right-hand side of these Eqs.\ are proportional to $m_Y n_Y$ \cite{erickcek_cannibalisms_2021}, but $m_Y n_Y = \rho_Y$ if the $Y$ particle has negligible kinetic energy.} We define the reheat temperature, $T_{\mathrm{RH}}$, by 
\begin{equation}
    \Gamma_Y \equiv \sqrt{\frac{8\pi}{3 m^2_{\text{pl}}} \left(\frac{\pi^2}{30} g_* T^4_{\mathrm{RH}} \right) } \,,\label{eq:T_RH}
\end{equation}
where $m_{\text{pl}}$ is the Planck mass and $g_*$ is the effective degrees of freedom of Standard Model relativistic particles. Since we are only considering reheat temperatures after BBN ($T_{\mathrm{RH}}<\SI{0.01}{MeV}$), $g_* = 3.38$. 

We assume neutrinos are composed of two massless species and one massive species with $m_{3} = 0.06$ eV. For simplicity, we also assume the abundance of $m_3$ does not change due to the $Y$ decay; while the injection of photons from the $Y$ particle does cause a non-standard abundance of $m_3$ (see Appendix \ref{sec:CLASSmodifications}), no new active neutrinos are produced from the $Y$ decay. Since we assume that the $Y$ particle decays after neutrino decoupling, any active neutrinos produced by the decay would lead to a non-thermal distribution, which would impact the matter power spectrum \citep{kawasaki_mev-scale_2000, cuoco_observations_2005, agarwal_effect_2011, das_non-thermal_2022}. However, the minimal assumption of $m_{3} = 0.06$ eV is well below cosmological bounds on the sum of neutrino masses \citep{planck_collaboration_planck_2020-1}. Therefore, while we assume no new active neutrinos are produced by the $Y$ decay, we do not expect relaxing this assumption to have a significant impact on our results. 

We include the two massless neutrinos in $\rho_{ur}$ and denote the energy density of the massive neutrino as $\rho_{ncdm}$ (non-cold dark matter). Since the massive neutrino is relativistic prior to recombination, the total radiation energy density is $\rho_r = \rho_\gamma + \rho_{ur} + \rho_{ncdm}$ for the reheat temperatures that we consider. 

We parametrize how much energy density the $Y$ particle contributes to the Universe via the maximum of $\rho_Y/\rho_r$, denoted as $\maxyr$. Under the assumption of radiation domination, this ratio can be expressed as
\begin{align}
    max\left(\frac{\rho_Y}{\rho_r}\right) = e^{\frac{1}{2}(\tilde{\Gamma}_Y-1)} \frac{\rho_{Y,i}}{\rho_{r,i}} \,\,\tilde{\Gamma}_Y^{-1/2}  \,, \label{eq:max_analitical}
\end{align}
where $ \rho_{x,i}\equiv \rho_x(a_i)$ for species $x$, $\widetilde{\Gamma}_Y \equiv \Gamma_Y/H_i$, $H_i \equiv H(a_i)$, and $a_i$ is the initial scale factor of our numerical solution (see Appendix \ref{sec:CLASSmodifications} for a  derivation).

\begin{figure}[h]
\centering
  \includegraphics[width=1\linewidth]{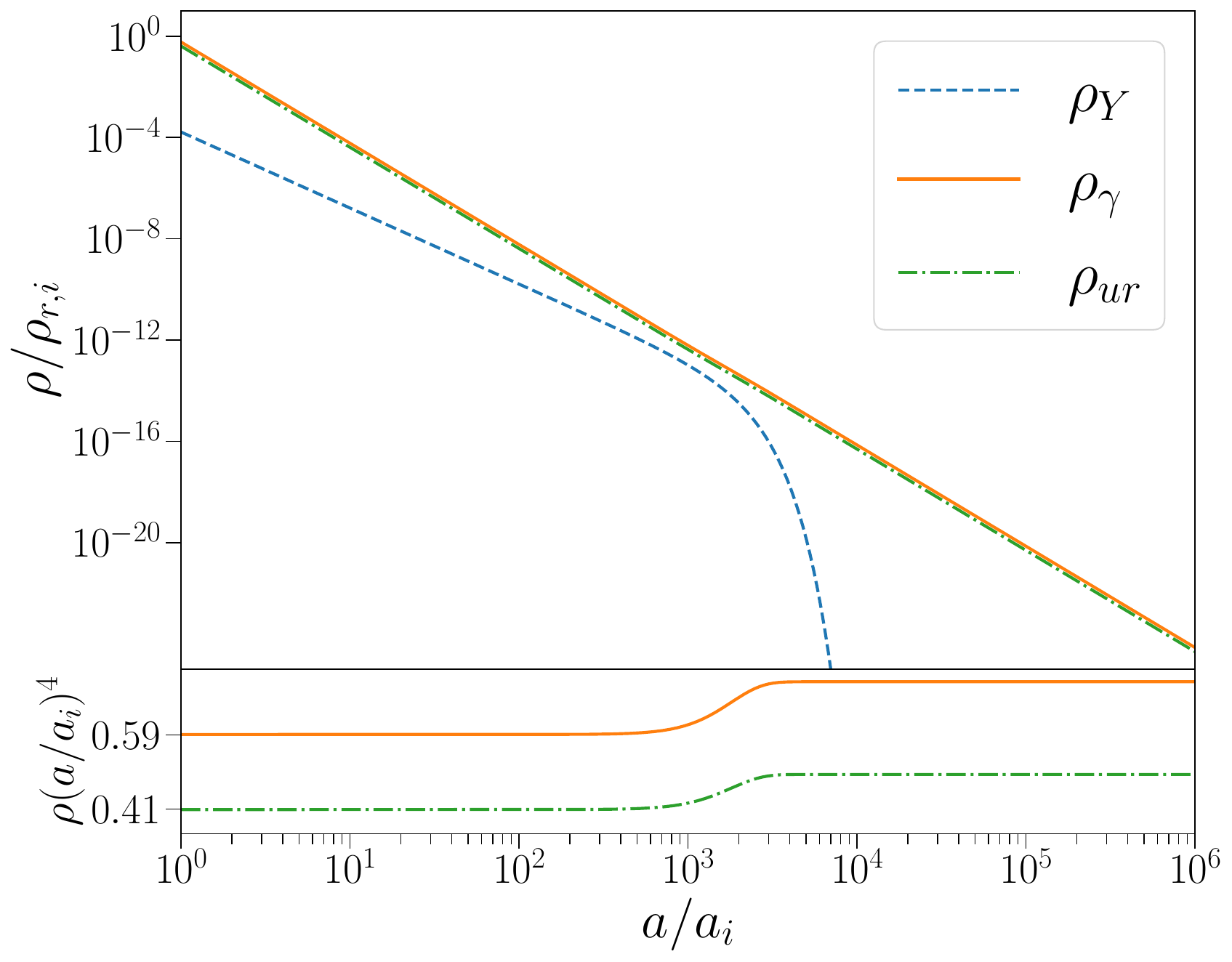}
  \caption{\footnotesize Numerical solution of Eqs.\ \eqref{eq:rho_Y}--\eqref{eq:rho_nu} for $\widetilde{\Gamma}_Y = 10^{-6}$, $f_\gamma = 0.6$, and $\maxyr = 0.1$. The $Y$ particle, taken to be non-relativistic, initially evolves as $\rho_Y \propto a^{-3}$ until the expansion rate equals the decay rate near $a/a_i\approx10^{3}$. Both $\rho_\gamma$ and $\rho_{ur}$ scale as $\rho \propto a^{-4}$ up until there is significant injection from the $Y$ decay at $a/a_i\approx10^{3}$.  Photons and other ultra-relativistic species gain 60\% and 40\% of the decay products, respectively. Then, $\rho_\gamma$ and $\rho_{ur}$ continue again as $a^{-4}$ after entropy injection. The comoving energy densities ($\rho a^4$) for photons and other relativisitc particles are plotted in the bottom panel to emphasize this evolution. }
  \label{fig:numerical}
\end{figure}

We set $\rho_{\gamma,i}$ such that the energy density of photons after $Y$-induced entropy injection corresponds to the fiducial $\Omega_{\gamma,0}$ determined by the measured CMB temperature of $T_0 = \SI{2.7255}{\kelvin}$. This initial condition for $\rho_\gamma$ is set by the parameter $\Omega'_{\gamma,0}$ such that the initial photon energy density is $\rho_{\gamma,i} = \Omega'_{\gamma,0} \rho_{\text{crit},0} \, a_i^{-4}$. The value of $\Omega'_{\gamma,0}$ can be calculated directly from our decay parameters $\maxyr$ and $f_\gamma$ (see Appendix \ref{sec:CLASSmodifications}). We choose $a_i$ to occur after electron-positron annihilation so the initial values for $\rho_{\gamma,i}$ and $\rho_{ur,i}$ are related by 
\begin{equation}
    \rho_{ur,i}+\rho_{ncdm,i} = \frac{7}{8} (3.044) \left(\frac{4}{11}\right)^{4/3} \rho_{\gamma,i} \,, \label{eq:neutrino_temps_bbn}
\end{equation}
where we enforce the effective number of neutrinos to be 3.044 pre-decay. The value of $\rho_{ncdm,i}$ is entirely determined by $\rho_{\gamma,i}$ (see Appendix \ref{sec:CLASSmodifications}), and the initial Hubble rate will be dominated by $\rho_{r,i} = \rho_{\gamma,i} + \rho_{ur,i} + \rho_{ncdm,i}$. From this, a value for $\rho_{Y,i}$ can be directly calculated from the parameters $\maxyr$ and $\Gamma_Y$ via Eq.\ \eqref{eq:max_analitical}. Therefore, we can fully describe the decay of the $Y$ particle with the parameters $\maxyr$, $f_\gamma$, and $\Gamma_Y$. Fig.\ \ref{fig:numerical} depicts a solution to Eqs.\ \eqref{eq:rho_Y}--\eqref{eq:rho_nu} for select values of $\widetilde{\Gamma}_Y$, $f_\gamma$, and $\maxyr$. 
 
\section{Effects of decaying particle} \label{sec:Effects}
The decaying $Y$ particle's imprint on the CMB anisotropy spectrum is primarily determined by two effects: the decay products change $\Neff$ at recombination and they change the primordial helium ($\YHe$) and deuterium (D/H) abundances by altering the expansion rate and baryon-to-photon ratio at BBN. In this section, we discuss these effects as well as distortions in the CMB frequency spectrum induced by the $Y$ decay. We compute the CMB temperature anisotropy power spectrum and spectral distortions with the Cosmic Linear Anisotropy Solving System (\textsc{CLASS}) \cite{blas_cosmic_2011}. For a discussion of the modifications made to CLASS during the implementation of our model, see Appendix \ref{sec:CLASSmodifications}.

\begin{figure}[t]
\centering
  \includegraphics[width=1\linewidth]{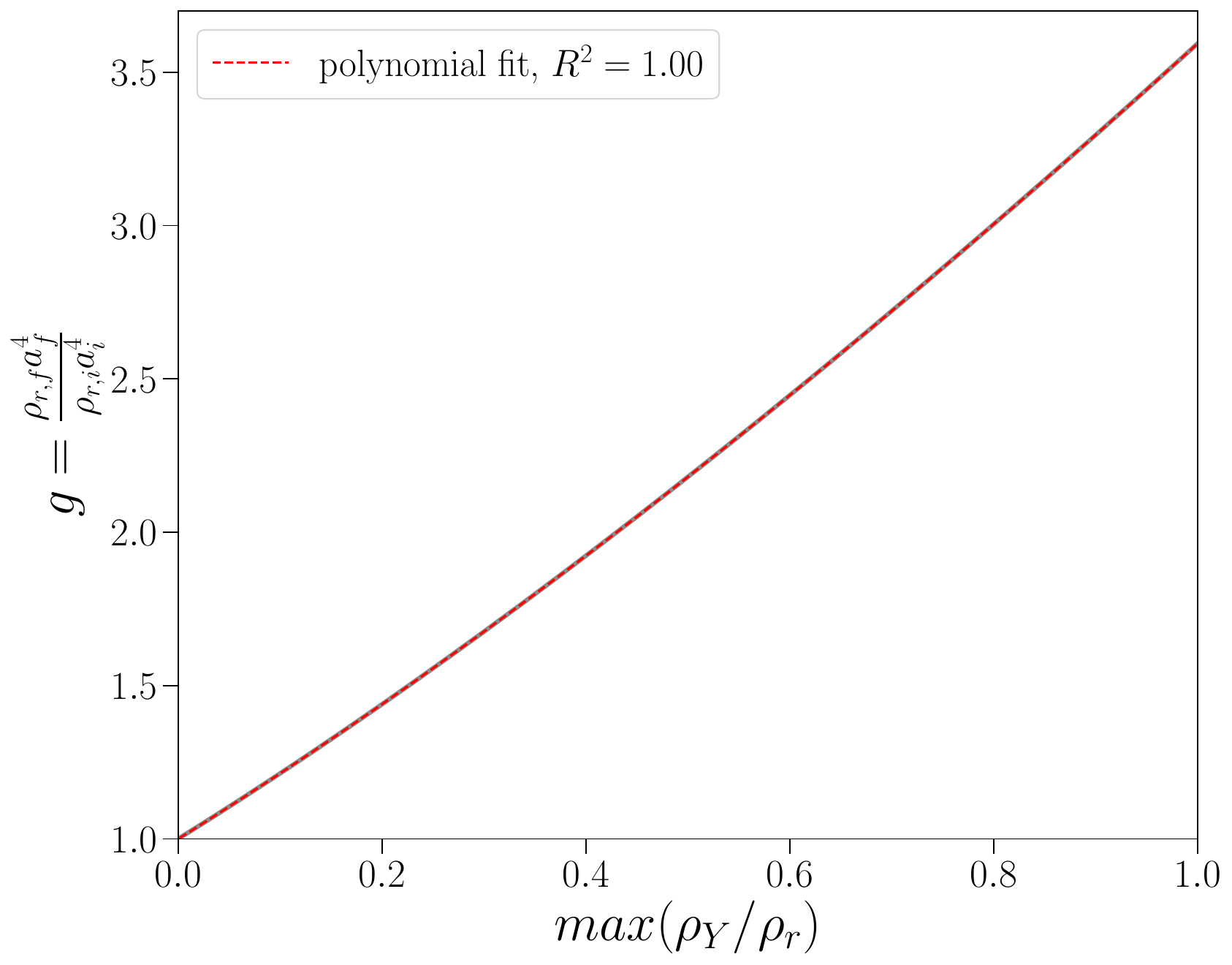}
  \caption{\footnotesize Relationship between $\maxyr$ defined by Eq.\ \eqref{eq:max_analitical} and the ratio of comoving radiation energy densities before and after $Y$ entropy injection (black line). This relationship is independent of the reheat temperature of the $Y$ particle. We fit this trend with a fourth order polynomial $g(x) = 0.0274x^4 -0.1525x^3 + 0.6458x^2 + 2.0727x + 1$, where $x=\maxyr$ for shorthand (red dashed line).}
  \label{fig:gap_maxYR}
\end{figure}

\subsection{Post-decay \texorpdfstring{$\Neff$}{TEXT}  } \label{sec:Neff}

\begin{figure*}[t]
  \includegraphics[width=1\linewidth]{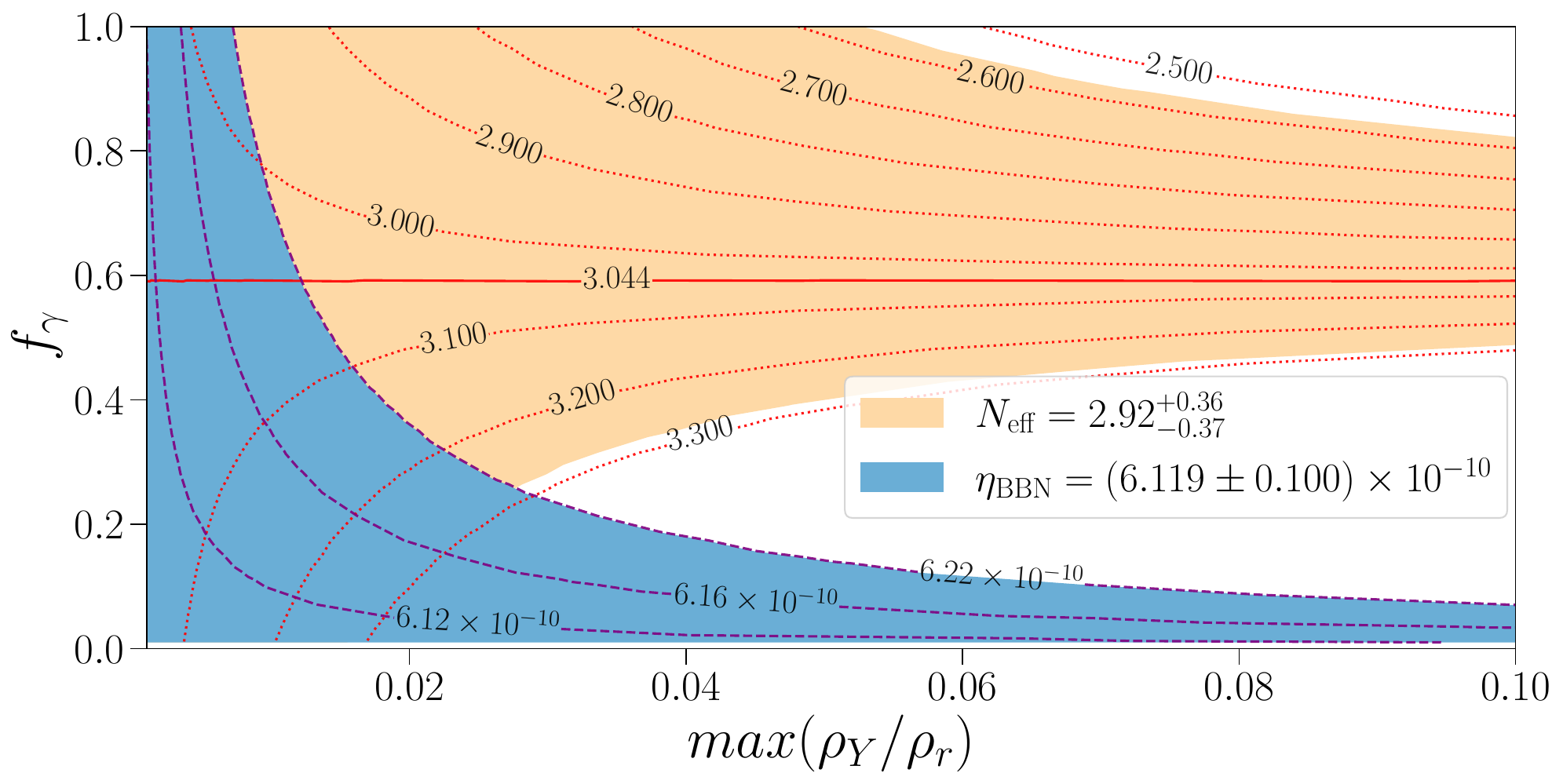}
  \caption{\footnotesize Post-decay $\Neff$ contours determined by $f_\gamma$ and $\maxyr$ via Eq.\ \eqref{eq:post-decay-Neff} are shown by the dotted red lines. The fiducial $\Neff$ value of $3.044$ is maintained as long as the photon fraction is $f_\gamma = 0.5913$ (solid red line). We include the 2018 \textit{Planck} TT,TE,EE+lowE 95\% confidence limits on $\Neff$ (orange shaded region) to illustrate the $Y$ particle parameter space that is naively consistent with CMB observations. Additionally, constraints on $\eta_{\text{BBN}}$ derived from observations of the deuterium abundance \cite{cooke_one_2018} are shown by the blue shaded region. The dashed purple lines show the values of $\eta_{\text{BBN}}$ calculated via Eq.\ \eqref{eq:eta_bbn}.}
  \label{fig:max_f_neff}
\end{figure*}

Since the decay of the $Y$ particle injects free-streaming radiation before recombination, the effective number of relativistic species, $\Neff$, will deviate from the fiducial value of 3.044. After electron-positron annihilation, entropy conservation dictates that the temperature ratio between  neutrinos and photons is $T_{\nu}/T_\gamma = (4/11)^{1/3}$. Entropy injection from the $Y$ decay will alter this temperature ratio by adding either photons or other relativistic species. We can parametrize this change in $T_{\nu}/T_\gamma$ as a change in $\Neff$ defined by
\begin{equation}
    \rho_{ur}+\rho_{ncdm} = \frac{7}{8} \Neff \left(\frac{4}{11}\right)^{4/3} \rho_{\gamma} \,. \label{eq:neutrino_temps}
\end{equation}
To relate the post-decay $\Neff$ to our model parameters, we first define 
\begin{equation}
g \equiv \frac{\rho_{r,f}a_f^{4}}{\rho_{r,i}a_i^{4}}\,,\label{eq:gap}
\end{equation}
where $\rho_{r,i}a_i^4$ and $\rho_{r,f}a_f^4$ are the comoving radiation energy densities before and after the $Y$-induced entropy injection, respectively. One would expect the change in comoving radiation energy density as a result of the $Y$ decay to be directly dependent on $\rho_Y$ when $H \sim \Gamma_Y$. Indeed, $g$ is a simple function of $\maxyr$. This can be seen in Fig.\ \ref{fig:gap_maxYR} where we numerically calculate $g$ for various values of $\maxyr$ defined by Eq.\ \eqref{eq:max_analitical} and fit the function with a fourth order polynomial.\footnote{If all of the $Y$ energy density was instantaneously transferred to $\rho_r$, then $g$ would equal $1+\maxyr$. However, as can be seen from Fig.\ \ref{fig:numerical}, most of the change in $\rho_r a^4$ occurs after $\rho_Y/\rho_r$ is maximized. As a result, $g-1 > \maxyr$. }

We can employ $g$ to determine the ratio of comoving energy densities before and after the $Y$ decay for the individual decay species:
\begin{flalign}
    g_\gamma\label{eq:gap_gamma} &\equiv \frac{\rho_{\gamma,f}a_f^4}{\rho_{\gamma,i}a_i^4}, &&\\\nonumber 
    &=  1 + f_\gamma\left(g-1\right) \left[1 + \frac{7}{8} (3.044) \left(\frac{4}{11} \right)^{4/3} \right];\\
    g_{ur} &\equiv \frac{\rho_{ur,f}a_f^4}{\rho_{ur,i}a_i^4}, &&\\\nonumber
    &= 1 + \frac{(1-f_\gamma)\left(g-1\right)}{N_{ur}} \left[ 3.044 + \frac{8}{7} \left( \frac{11}{4}\right)^{4/3} \right] \,\,.
\end{flalign}
Here, $N_{ur} = 2.0308$ is the effective number of ultra-relativistic species before the $Y$ decay, excluding the single massive neutrino (see Appendix \ref{sec:CLASSmodifications}). 

The single massive neutrino species is still relativistic at recombination and contributes to $\Neff$. As shown in Appendix \ref{sec:postdecay}, it follows that the post-decay $\Neff$ is 
\begin{equation}
    \Neff = \frac{1}{g_\gamma} \left[ 3.044 + N_{ur}(g_{ur} - 1) \right] \,.\label{eq:post-decay-Neff}
\end{equation}
This relationship is shown in Fig.\ \ref{fig:max_f_neff} where we plot contours of the post-decay $\Neff$ for a range of decay scenarios described by $f_\gamma$ and $\maxyr$ (red dotted lines). Most notably, there is a ``sweet spot" fraction of $f_\gamma \simeq 0.59$ that maintains $\Neff = 3.044$ after the decay for all values of $\maxyr$ (shown by the solid red line). Fig.\ \ref{fig:max_f_neff} shows that $f_\gamma < 0.59$ corresponds to an increase in $\Neff$ since the decay of the $Y$ particle increases the relative energy density of ultra-relativistic species.
Conversely, $f_\gamma > 0.59$ decreases $\rho_{ur}/\rho_\gamma$ and thereby results in a decrease in $\Neff$.

Some combinations of $f_\gamma$ and $\maxyr$ will not be favored by \textit{Planck} data. This is demonstrated by the orange contour in Fig.\ \ref{fig:max_f_neff}, showing the 2018 \textit{Planck} TT,TE,EE+lowE reported bounds on $\Neff$ \cite{planck_collaboration_planck_2020}. For scenarios in which $f_\gamma$ deviates from the ``sweet spot" value, \textit{Planck} $\Neff$ bounds require smaller values of $\maxyr$. However, it is important to note that naively applying these \textit{Planck} $\Neff$ bounds to our model parameters would be  incorrect, as doing so would ignore the fact that there is a degeneracy between $\Neff$ and the primordial helium abundance, $\YHe$, which is also affected by the $Y$ particle and its decay products. 

To understand this degeneracy with $\YHe$, we must first understand how changes in $\Neff$ affect the CMB anisotropy spectrum. Increasing $\Neff$ increases the pre-recombination expansion rate and therefore decreases the sound horizon $r_s \propto 1/H$. In order to maintain the precisely measured angular size of the sound horizon, $\theta_s$, one must increase $H_0$ in order to decrease the angular diameter distance to the CMB such that $\theta_s$ is fixed. Meanwhile, the photon diffusion length scales as $r_D \propto \sqrt{1/H}$ and so increasing $\Neff$ while simultaneously keeping $\theta_s$ fixed leads to an increase in the angular size of the diffusion length, $\theta_D$ \cite{hou_how_2013,planck_collaboration_planck_2014}. In summary, an increase in $\Neff$ results in more Silk damping on small angular scales (and vice versa). 

\begin{figure}[t]
\centering
  \includegraphics[width=1\linewidth]{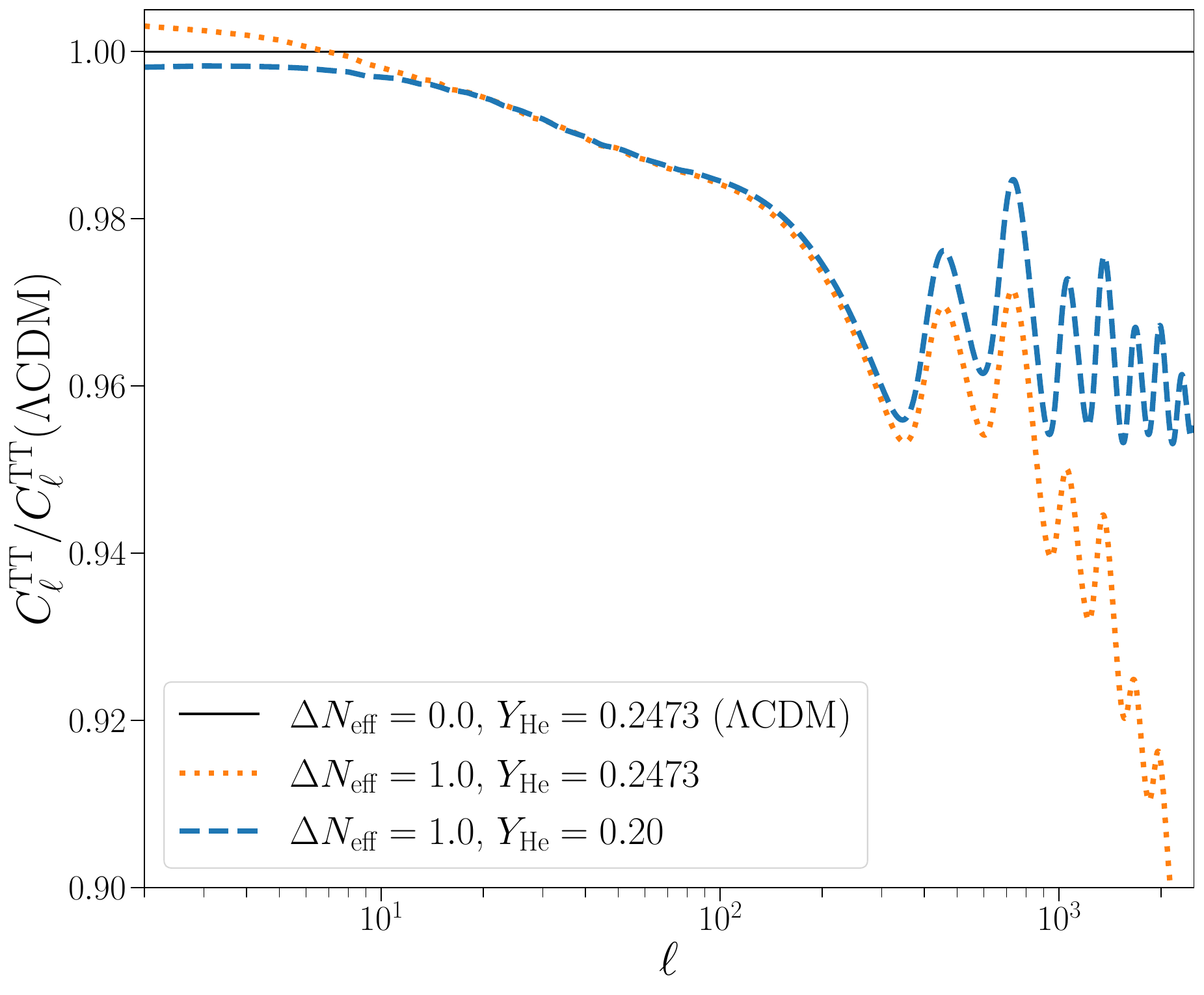}
  \caption{\footnotesize Effects on the CMB temperature power spectrum from varying $\Neff$ and $\YHe$ at fixed $\omega_b$, $z_{eq}$, and $\theta_s$. Increasing $\Neff$ compared to $\Lambda$CDM results in more damping on small scales (orange dotted line). This effect can be mitigated by decreasing $\YHe$ (blue dashed line). The beat frequency seen in these residuals is a direct consequence of the free-streaming nature of neutrinos producing a unique phase shift in the CMB acoustic peaks.}
  \label{fig:neff_damping}
\end{figure}

Altering the helium abundance can mitigate the effects of changing $\Neff$ on the dampening tail. Decreasing $\YHe$ results in more free electrons at recombination, which decreases the Compton mean free path, resulting in less photon diffusion and thus less damping of small-scale anisotropies. This behavior is depicted in Fig.\ \ref{fig:neff_damping}. The dotted orange line shows an increase in $\Neff$ resulting in more damping on small scales compared to $\Lambda$CDM (for fixed $\omega_b$, $z_{eq}$, and $\theta_s$, where $\omega_b\equiv\Omega_{b,0}h^2$ is the present-day baryon energy density and $z_{eq}$ is the redshift of matter-radiation equality). Simultaneously decreasing $\YHe$ as $\Neff$ increases, shown by the blue dashed line, can mitigate the damping. The phase shift observed in Fig.\ \ref{fig:neff_damping} is a lingering effect of changing $\Neff$; free-streaming relativistic particles generate a unique phase shift in the acoustic peaks \cite{bashinsky_signatures_2004,follin_first_2015,baumann_phases_2016}.

Therefore, even though our model parameters $f_\gamma$ and $\maxyr$ conspire to produce a non-zero $\Delta \Neff$, altering $\YHe$ in a inverse manner (i.e. raising $\YHe$ as $\Neff$ decreases, or vice versa) can mitigate the effects of Silk damping that a changing $\Neff$ causes. Interestingly, our model already necessitates an increase in $\YHe$ by altering the expansion rate and baryon-to-photon ratio at BBN.

\subsection{Change in primordial abundances}\label{sec:YHe}
Since the injection of photons from the $Y$ decay necessitates a decrease in $\rho_\gamma$ pre-decay, any $f_\gamma \neq 0$ decay alters the baryon-to-photon ratio at the time of BBN, $\eta_{\text{BBN}}$. If $\eta_0$ is the baryon-to-photon ratio at recombination inferred by $\eta_0 = 6.0 \times 10^{-10} \left(\omega_b/0.022 \right)$, then the increased ratio prior to the decay of the $Y$ particle is
\begin{align}
    \eta_{\text{BBN}} =g_\gamma^{3/4} \eta_0 \,. \label{eq:eta_bbn}
\end{align}
Given that $g_\gamma$ is completely determined by $f_\gamma$ and $\maxyr$ via Eq.\ \eqref{eq:gap_gamma}, we are able to place constraints on our decay parameters via constraints on $\eta_{\text{BBN}}$. This relationship is shown in Fig.\ \ref{fig:max_f_neff} where we calculate $\eta_{\text{BBN}}$ as a function of $f_\gamma$ and $\maxyr$ for a fixed value of $\omega_b = 0.02238$ (dashed purple lines). The blue contour shows the parameter space that is allowed by the bounds $\eta_{\text{BBN}} = (6.119 \pm 0.100)\times10^{-10}$ derived from observations of the primordial deuterium abundance \cite{cooke_one_2018}. Since large $f_\gamma$ values correspond to significant photon injection, any increase in $\maxyr$ at a large $f_\gamma$ leads to a significant change in $\etaBBN$. On the other hand, we can see that large values of $\maxyr$ are allowed by bounds on $\etaBBN$ as $f_\gamma\rightarrow0$ since fewer photons are being injected by the decay and so $g_\gamma$ approaches unity (i.e. $\eta_{\text{BBN}}\rightarrow \eta_0$).

Much like the bounds on $\Neff$ discussed in Sec.\ \ref{sec:Neff}, these bounds on $\etaBBN$ seen in Fig.\ \ref{fig:max_f_neff} are only meant to depict an approximation for the allowed parameter space; the true bounds will be subject to uncertainties in $\omega_b$, which are taken into account later. Furthermore, the inference of $\etaBBN$ from deuterium assumes a standard expansion history during BBN. Decay scenarios with large $\maxyr$ and short lifetimes that end soon after the end of BBN ($\tau_Y \approx 10^{4}$ sec) will contribute a non-negligible energy density \textit{during} BBN. 

We therefore altered the BBN code PArthENoPe v3.0 \cite{gariazzo_parthenope_2022} to include the contribution of $\rho_Y$ to the expansion rate during BBN. We modified PArthENoPe to accept a new input parameter, $\rho_Y^{\mathrm{BBN}}$, which is the value of $\rho_Y$ at a reference temperature of $T=0.1$ MeV, and we assume the $Y$ particle is non-relativistic at all temperatures less than $\SI{1}{MeV}$ so that $\rho_Y \propto a^{-3}$. This $\rho_Y^{\mathrm{BBN}}$ parameter is entirely determined by our decay parameters $\maxyr$ and $\Gamma_Y$. Using our modified version of PArthENoPe, we created lookup tables for CLASS that read in $\etaBBN$ and $\rho_Y^{\mathrm{BBN}}$ and produce values for the helium and deuterium abundances. These tables were created with $\Delta\Neff = 0$ (our initial conditions ensure that $\Neff = 3.044$ at BBN) and a neutron lifetime of $\tau = \SI{879.4}{\second}$.

\begin{figure}[h]
\centering
  \includegraphics[width=1\linewidth]{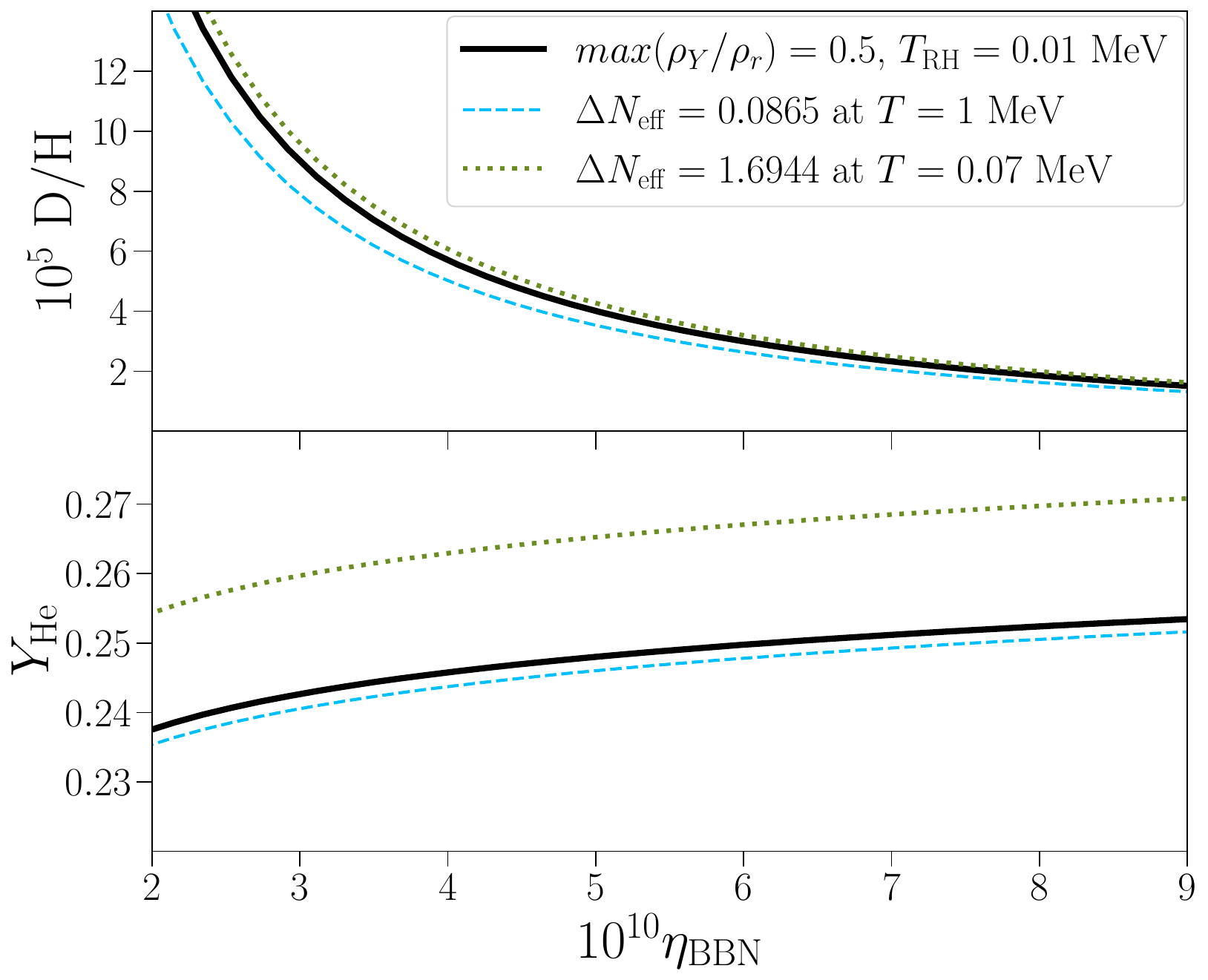}
  \caption{\footnotesize Dependence of primordial abundances on $\etaBBN$ for a decay scenario that decays right after the end of BBN (black solid line). Matching the non-standard Hubble rate during BBN as a result of $\rho_Y$ with a constant nonzero $\Delta\Neff$ does not reproduce the same abundances. Matching Hubble at $T=1$ MeV when neutron-proton freeze-out occurs (blue dashed line) is a good approximation for $\YHe$, but underestimates D/H. Matching Hubble at $T=0.07$ MeV when helium production freezes out (green dotted line) is a good approximation for D/H, but overestimates $\YHe$. }
  \label{fig:bbn_abundances}
\end{figure}

Since the $Y$ particle is non-relativistic during BBN and evolves as $\rho_Y\propto a^{-3}$, its contribution to $H$ cannot be modeled with a constant $\Delta \Neff$. Instead, one can think of the contribution of $\rho_Y$ to the expansion rate during BBN as an evolving $\Delta \Neff$; at any given temperature, there is a nonzero $\Delta \Neff$ that matches the non-standard Hubble rate due to the inclusion of $\rho_Y$ \cite{hufnagel_bbn_2018-1}. This behavior is demonstrated in Fig.\ \ref{fig:bbn_abundances}. The black solid lines track the abundances of deuterium (D/H) and helium ($\YHe$) as a function of $\etaBBN$ for a specific decay scenario in which the $Y$ particle decays away right after the end of BBN ($T_{\mathrm{RH}} = 0.01$ MeV). If we then match the Hubble rate at a temperature of $T=1$ MeV using a $\Delta\Neff = 0.0865$ (blue dashed line), then we mostly recover the correct $\YHe$ since the abundance of helium is primarily set by neutron-proton freeze-out near $T=1$ MeV. However, the abundance of deuterium is set when helium production freezes out around $T = 0.07$ MeV. So, matching at $T=1$ MeV correctly approximates $\YHe$, but fails to yield the correct D/H. In a similar manner, matching at $T=0.07$ MeV with $\Delta\Neff = 1.6944$ (green dotted line) results in a good approximation for the deuterium abundance, but grossly overestimates the correct abundance of helium. Therefore, the contribution of $\rho_Y$ to the Hubble rate during BBN cannot be captured by a simple constant change to $\Neff$. 

The abundances of deuterium and helium are both sensitive to $\etaBBN$ and the expansion rate. Increasing the expansion rate during BBN leads to an increase in both $\YHe$ and D/H, while increasing $\etaBBN$ results in larger $\YHe$ and smaller D/H. Any $Y$ decay scenario in which the decay products consist of photons sees an increase in both $\eta_{\text{BBN}}$ and $H$, and thus an increase in $\YHe$. On the other hand, the abundance of deuterium is predominately sensitive to changes in $\etaBBN$. So, even though $\rho_Y$ increases the expansion rate during BBN, we expect an overall decrease in D/H for any $f_\gamma \neq 0 $ decay scenario. 

Finally, we note that all $f_\gamma \neq 0$ scenarios result in a larger $\YHe$ compared to $\Lambda$CDM while changes in $\Neff$ are symmetric about $f_\gamma \approx 0.59$ (see Fig.\ \ref{fig:max_f_neff}). Therefore, we expect an asymmetry in the small-scale Silk damping resulting from decays of different $f_\gamma$. Decays with $f_\gamma > 0.59$ will have a lack of small scale damping from a decreased $\Neff$, but an increased $\YHe$ compensates by enhancing Silk damping. On the other hand, scenarios with $f_\gamma < 0.59$ will have combined damping effects of both an increased $\YHe$ and $\Neff$. This asymmetry suggests a preference for decay scenarios with $f_\gamma > 0.59$.

\subsection{Spectral Distortions}\label{sec:distortion calculations}
Deviations from the blackbody spectrum of the CMB, known as spectral distortions (SDs), are the result of energy injection into the photon bath of the CMB at a late enough time such that the photons do not thermalize before recombination \cite{zeldovich_interaction_1969, sunyaev_interaction_1970, illarionov_comptonization_1975}. The form of these SDs depends on the time of energy injection; the era of $\mu$ distortions occurs roughly in the redshift range of $5\times10^4 < z < 2\times10^6$, while $y$-type distortions result from injection at $z<5\times10^4$ \cite{chluba_evolution_2012,lucca_synergy_2020}. For $z>2\times10^6$, thermalization processes are very efficient and so any energy injection results in a temperature shift to the observed blackbody. Since we will be considering reheat temperatures that correspond to a range of about $10^5 < z < 4.5\times10^7$, $\mu$ distortions will have the most constraining power.

We calculate the distortions in \textsc{CLASS} using a new SD module based on the work of \textcite{lucca_synergy_2020}. This module calculates $\mu$ and $y$ distortions by
\begin{align}
    \tilde{d} = \frac{\Delta \rho_\gamma}{\rho_\gamma}\Bigg|_d &\equiv \int \frac{dQ/dz}{\rho_\gamma} \cdot \mathcal{J}_d(z) dz \,, \label{eq:distortion}
\end{align}
 where $d$ indexes the distortion type,\footnote{\textcite{lucca_synergy_2020} uses ``$a$" to index distortion type, but we choose to use ``$d$" in order to avoid confusion with scale factor.} $dQ/dz$ is the energy injection rate, and $\mathcal{J}_d(z)$ is a branching ratio that dictates the contribution of an energy injection to a specific distortion type. The energy injection rate can be recast as a rate with respect to proper time, $t$, via
\begin{align}
    \frac{dQ/dz}{\rho_\gamma} &= -\frac{\dot{Q}}{(1+z)H\rho_\gamma} \,,
\end{align}
and we define the energy injection rate of our decaying model as
\begin{align}
    \dot{Q}(t) = \rho_{Y}(t) f_\gamma \Gamma_Y e^{-\Gamma_Y t} \,.
\end{align}
The distortion parameters are then determined by Eq.\ \eqref{eq:distortion} and setting $\mu = 1.4 \tilde{\mu}$ and $y =\tilde{y}/4$. 

\begin{figure}[t]
\centering
  \includegraphics[width=1\linewidth]{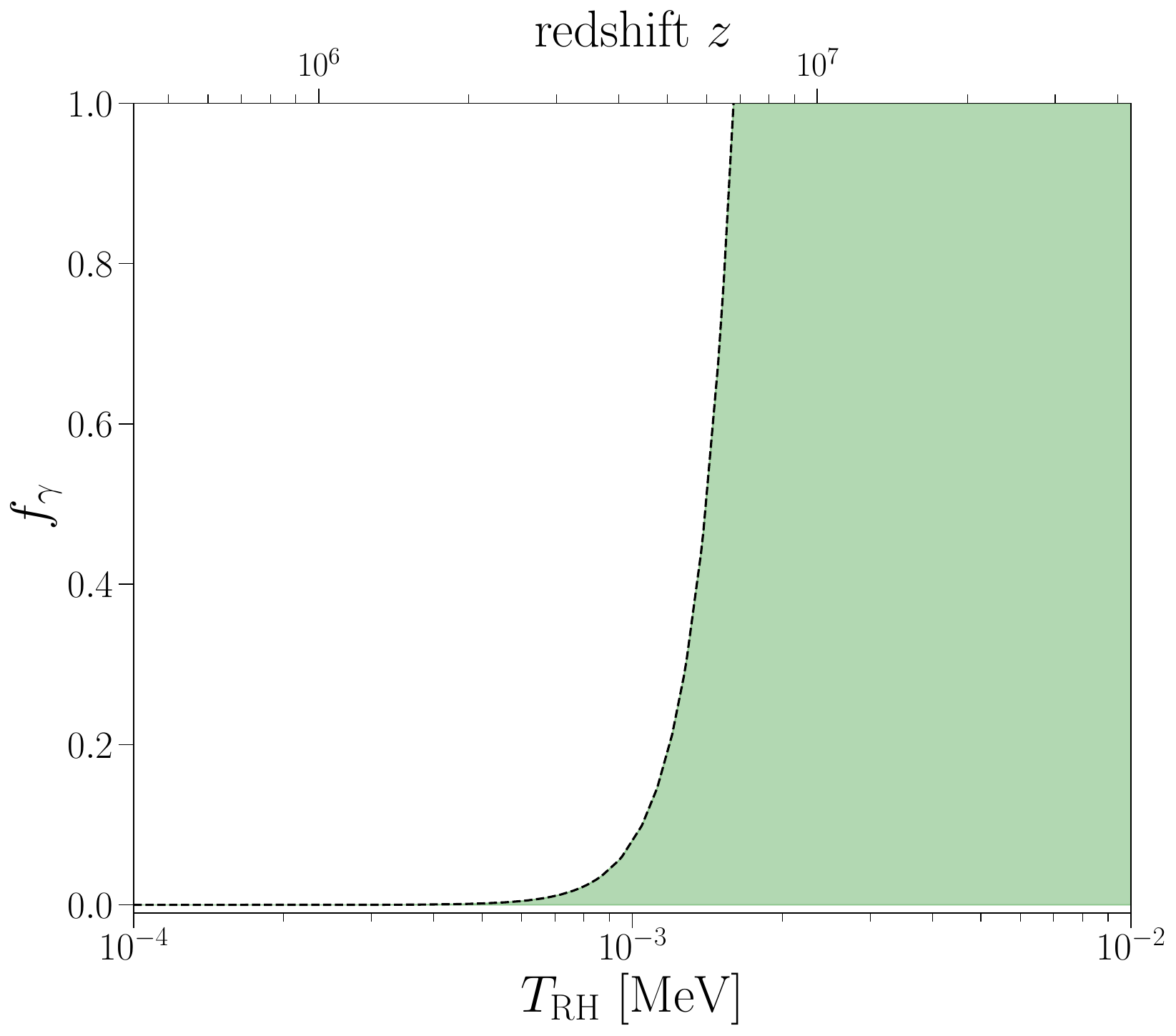}
  \caption{\footnotesize Example of expected $\mu$ distortion constraints for a value of $\maxyr = 0.01$. The green shaded region shows the parameter space allowed by the COBE/FIRAS constraint $|\mu| < 9\times10^{-5}$. Reheat temperatures cooler than $T_{\mathrm{RH}} \approx 9\times10^{-4}$ MeV are only allowed by SDs if $f_\gamma \rightarrow 0$.  }
  \label{fig:mu_constraints}
\end{figure}

The COBE/FIRAS satellite measured the blackbody spectrum of the CMB and determined upper bounds on the distortions to be $|\mu| < 9\times10^{-5}$ and $|y|<1.5\times10^{-5}$ (95\% C.L.) \cite{mather_measurement_1994,fixsen_cosmic_1996}. In Fig.\ \ref{fig:mu_constraints}, we calculate $|\mu|$ for a range of $f_\gamma$ and $T_{\mathrm{RH}}$ values using \textsc{CLASS} and depict the parameter space allowed by the COBE/FIRAS bound by the green shaded region. Here we see that the COBE/FIRAS bound on $\mu$ allows photon injection for reheat temperatures hotter than about $9\times10^{-4}$ MeV, but requires that $f_\gamma$ quickly approaches zero for $T_{\mathrm{RH}}\lesssim 9\times10^{-4}$ MeV ($\Gamma_Y \lesssim \SI{3.08e-7}{\per\second}$ ). 

Since the CMB angular scales that are accessible to \textit{Planck} ($\ell \lesssim 2500$) are dominated by modes that enter the horizon at a temperatures below $2\times10^{-5}$ MeV, we can neglect perturbations related to the $Y$ particle for cases in which $f_\gamma \gtrsim 0.01$ because those values will only be allowed by SDs when $T_{\mathrm{RH}} \gtrsim 9\times10^{-4}$ MeV. However, scenarios with $f_\gamma=0$ are not restricted in reheat temperature by SDs, and therefore require consideration of the $Y$ perturbation Eqs.\ as well as corrections to the ultra-relativistic species perturbations. We reserve the inclusion of these perturbations for a subsequent analysis and, in this work, focus solely on $f_\gamma \geq 0.01$. Nevertheless, our results can be extended to smaller values of $f_\gamma$, provided that $T_{\mathrm{RH}} > 9.5\times10^{-4}$ MeV \mbox{($\Gamma_Y > \SI{3.43e-7}{\per\second}$ )}.  For these decay rates, the posteriors for all parameters are the same when comparing $f_\gamma$ fixed at either $f_\gamma = 0.01$ or $f_\gamma = 0$ (see Fig.\ \ref{fig:fzero_f0p01} in Appendix \ref{sec:MCMCruns}). 

\section{Analysis Method} \label{sec:MCMC}
We derive constraints for $\Gamma_Y$, $\maxyr$, and $f_\gamma$ by conducting a MCMC analysis  using \textsc{MontePython-v3} \cite{audren_conservative_2013,brinckmann_montepython_2018} with our modified version of \textsc{CLASS}. This analysis is implemented with a Metropolis-Hastings algorithm and flat priors on the base six cosmological parameters \{$\omega_b$, $\omega_{cdm}$, $\theta_s$, $A_s$, $n_s$, $\tau_{reio}$\}. We assume neutrinos are composed of two massless and one massive species with $m_{3} = 0.06$ eV. In order to account for the fact that the $Y$ particle decaying into photons changes $T_\nu/T_\gamma$, we provide CLASS with a corrected temperature of the massive neutrino (see Appendix \ref{sec:CLASSmodifications}). We employ \textit{Planck} high-$\ell$ TTTEEE, low-$\ell$ TT, and low-$\ell$ EE likelihood functions \cite{planck_collaboration_planck_2020-1} and refer to the combination of these data as \textbf{Planck}. 

As discussed in Sec.\ \ref{sec:YHe}, the $Y$ decay affects the abundances of $\YHe$ and D/H through its influence on $\etaBBN$ and the expansion rate during BBN. Current measurements of the deuterium abundance have reached a $\sim1\%$ precision level with $(\mathrm{D/H}) = (2.527\pm0.030)\times10^{-5}$ \cite{cooke_one_2018}. However, we note that this bound only includes measurement uncertainty (i.e.\ $\sigma_{\mathrm{MEAS}} = 3.0 \times 10^{-7}$). \textcite{cooke_one_2018} cite uncertainties associated with BBN calculations for two different values of the $d(p, \gamma)^3\text{He}$ cross-section; using the computationally inferred cross-section leads to a $2\sigma$ discrepancy with CMB measurements of $\etaBBN$, while utilizing the measured cross-section suggested by \textcite{adelberger_solar_2011} yields $\etaBBN = (6.119\pm0.100)\times10^{-10}$. \textcite{millea_new_2015} fold in uncertainty associated with nuclear reaction rates to D/H by assuming $\sigma_{\mathrm{NUCL}} = 4.5\times10^{-7}$ and adding uncertainties in quadrature such that $\sigma_{\mathrm{DH}} = 5.4\times10^{-7}$. It is unclear which $d(p, \gamma)^3\text{He}$ cross-section is used for this uncertainty, and \textcite{millea_new_2015} assume a linear relationship between D/H and $\eta$. Therefore, we translate the $\etaBBN = (6.119\pm0.100)\times10^{-10}$ bounds reported by \textcite{cooke_one_2018} into an uncertainty in D/H. We calculate D/H for a range of $\etaBBN$ with our modified version of PArthENoPe for $\Delta\Neff^{\mathrm{BBN}}=0$ and a neutron lifetime of $\SI{879.4}{\sec}$. We find that $\text{D/H}\propto \etaBBN^{-1.65}$ for the range of $\etaBBN$ relevant to this work and that this fit is relatively insensitive to small changes in $\Neff^{\mathrm{BBN}}$. We use this power law to infer a new fractional uncertainty on D/H ($\sigma_{\text{DH}}/{\text{DH}} = 1.65\times \sigma_{\etaBBN}/{\etaBBN}$). Doing so results in $(\text{D/H}) = (2.527 \pm 0.068)\times10^{-5}$. We create a Gaussian likelihood function in \textsc{MontePython} for D/H for which the mean and standard deviation are set to be $\mu_{\mathrm{DH}} = 2.527\times10^{-5}$ and $\sigma_{\mathrm{DH}} = 6.83\times10^{-7}$, respectively. We denote this Gaussian likelihood as \textbf{D/H}.

Finally, we calculate the spectral distortions $\mu$ and $y$ in CLASS according to Sec.\ \ref{sec:distortion calculations}. Similar to our constraint on D/H, we add a Gaussian likelihood for both $\mu$ and $y$. These Gaussian likelihood functions were taken to have a mean of zero and a 2$\sigma$ deviation equal to the respective upper bounds $|\mu| < 9\times10^{-5}$ and $|y|<1.5\times10^{-5}$ determined by COBE/FIRAS. We refer to the combination of these $\mu$ and $y$ likelihoods as \textbf{SD}.

We used the Gelman-Rubin \cite{gelman_inference_1992} criterion $|R-1| < 0.04$ to asses convergence of our MCMC chains.\footnote{All $\Lambda$CDM runs have $|R-1| < 0.01$. This bound of $0.04$ was selected due to the non-Gaussian nature of the posteriors of our decay parameters.} Post-processing of chains was done using \textsc{GetDist} \cite{lewis_getdist_2019} and removing the first 30\% of points as burn-in.

\subsection{Initial priors} \label{sec:priors}
Figure \ref{fig:max_f_neff} demonstrates that naive bounds from $\Neff$ would leave $\maxyr$ unconstrained if $f_\gamma$ is such that $\Neff=3.044$ after the $Y$ decay. The addition of $\etaBBN$ bounds, however, suggests that $\maxyr$ would be limited to less than 0.05. Therefore, we choose to sample $\maxyr = [0, 0.07]$ when including the D/H likelihood. Otherwise, when only considering Planck or Planck+SD, we sample $\maxyr = [0, 0.8]$ in order to sufficiently explore the asymptotic behavior around the ``sweet spot'' $f_\gamma$ shown in Fig.\ \ref{fig:max_f_neff}. A value of $\maxyr=0$ corresponds to standard $\Lambda$CDM.

As discussed in Sec.\ \ref{sec:distortion calculations}, the smallest scale accessible to \textit{Planck} enters the horizon at a temperature of about $2\times10^{-5}$ MeV. Therefore, a full perturbation analysis of the $Y$ particle and its decay products is required for reheat temperatures $T_{\mathrm{RH}} \lesssim 2\times10^{-5}$ MeV. We retain this for a subsequent analysis and instead restrict our current work to $T_{\mathrm{RH}} > 2\times10^{-5}$ MeV.
Furthermore, CLASS begins evolving a given perturbation mode outside the cosmological horizon with adiabatic initial conditions. In order to avoid the $Y$ decay influencing these initial conditions, we choose to only consider $T_{\mathrm{RH}} \geq 9.5\times10^{-4}$ MeV so that $Y$-induced entropy injection is complete by the time CLASS begins evolving perturbation modes relevant to \textit{Planck} ($k \lesssim \SI{0.18}{\per\mega\parsec}$ or $\ell \lesssim 2500$). Assuming BBN has completed by $T \approx 0.01$ MeV, this restriction on $T_{\mathrm{RH}}$ leads to a prior of  $T_{\mathrm{RH}} = [9.5\times10^{-4}, 10^{-2}]$ MeV ( $\Gamma_Y = [3.43\times10^{-7}, 3.80\times10^{-5}] \,\SI{}{\per\second} $). 

Spectral distortions will confine $f_\gamma$ to be very small as $T_{\mathrm{RH}}$ approaches $9.5\times10^{-4}$ MeV. Therefore we would like to sample small values of $f_\gamma$ for colder reheat temperatures, but explore $f_\gamma$ of order unity when $T_{\mathrm{RH}} \gg 9.5\times10^{-4}$ MeV. A flat prior of $f_\gamma=[0,1]$ fails to sufficiently sample small values of $f_\gamma$ as $T_{\mathrm{RH}} \rightarrow 9.5\times10^{-4}$ MeV, but using a flat prior on $\log_{10}f_\gamma$ does not properly depict the parameter space of large $f_\gamma$. Furthermore, small values of $f_\gamma$, such as $f_\gamma = 10^{-5}$, would be indicative of a finely tuned decay scenario. To avoid these scenarios and achieve sufficient sampling, we choose to sample $f_\gamma = [0.01, 1]$. The initial priors used in our analysis are summarized in Table \ref{tab:initial_priors}. 

\begin{table}[t]
\caption{\footnotesize Summary of priors used for decay parameters with each likelihood combination.}
    \centering
\begin{tabular}{ccc}
    \hline
    \hline
    & \multicolumn{1}{l}{\textbf{Planck}, \textbf{Planck}+\textbf{SD}\,\, \, } & \multicolumn{1}{r}{\,\,\,\textbf{Planck}+\textbf{SD}+\textbf{D/H}} \\  
    $f_\gamma$ & \multicolumn{2}{c}{[0.01, 1] } \\
    $T_{\mathrm{RH}}$ [MeV] & \multicolumn{2}{c}{$[9.5\times10^{-4}, 10^{-2}]$} \\
    $\maxyr$ & [0, 0.8] & [0, 0.07] \\
    \hline
    \hline
\end{tabular}
    \label{tab:initial_priors}
\end{table}

\section{Results} \label{sec:results}
\begin{figure*}[t]
\centering
  \includegraphics[width=\linewidth]{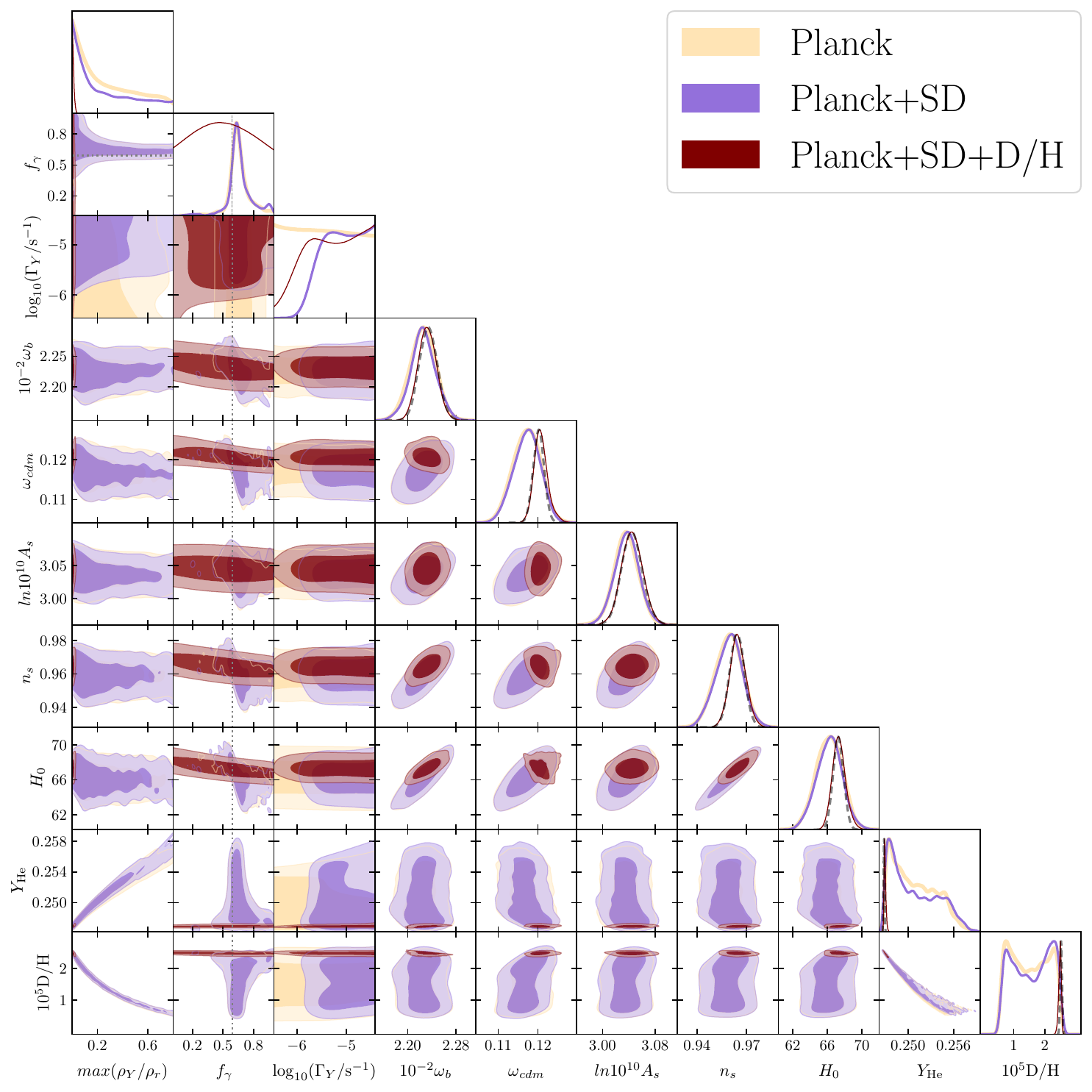}
  \caption{\footnotesize 1D and 2D posterior distributions of decay and cosmological parameters for different combinations of \textit{Planck} high-$\ell$ TT,TE,EE, low-$\ell$ TT, and low-$\ell$ EE  (\textbf{Planck}) data, CMB spectral distortions (\textbf{SD}) bounds, and bounds on the observed deuterium abundance (\textbf{D/H}). We include the 1D posteriors for $\Lambda$CDM constrained by Planck (dashed black line). The dotted gray line traces $f_\gamma = 0.5913$, which maintains $\Neff = 3.044$ at recombination.}
  \label{fig:triangle}
\end{figure*}

Figure \ref{fig:triangle} shows the resulting posterior distributions of the decay parameters as well as $\omega_b$, $\omega_{cdm}$, $A_s$, $n_s$, $H_0$, $\YHe$, and D/H, with 68\% and 95\% CL contours for Planck, Planck+SD, and Planck+SD+D/H datasets. Table \ref{tab:posteriors} summarizes the mean and $2\sigma$ errors for each parameter. For the posteriors of all base six cosmological parameters, see Fig.\ \ref{fig:triangle_full} in Appendix \ref{sec:MCMCruns}.

In Fig.\ \ref{fig:triangle}, it can be seen that the Planck data favors changes in $\omega_b$, $\omega_{cdm}$, $A_s$, and $H_0$ in order to accommodate a $Y$ decay scenario. For cases in which $f_\gamma$ is greater than the ``sweet spot" value of $0.5913$, $\Neff$ is less than $3.044$, which results in a larger $z_{eq}$ compared to $\Lambda$CDM. The Planck data, being very sensitive to $z_{eq}$ via the early ISW effect, favors a decrease in $\omega_{cdm}$ in order to restore $z_{eq}$ to its fiducial value. Furthermore, the $Y$ induced entropy injection directly alters the pre-recombination expansion history and therefore the size of the sound horizon, $r_s$. In the case of $f_\gamma > 0.5913$, $r_s$ is increased compared to $\Lambda$CDM. The decrease in $\omega_{cdm}$ that is required to maintain a fixed $z_{eq}$ results in a larger angular diameter distance, $d_A$. However, the rate of change for $r_s$ is greater than that of $d_A$ and thus leads to an increase in the angular size of the sound horizon $\theta_s$. Since $\theta_s$ sets the anisotropy peak locations, the Planck data tend to reduce this increase in $\theta_s$ by decreasing $H_0$.

\begin{table*}[t]
\centering
    \caption{\footnotesize Mean and $2\sigma$ errors on decay and cosmological parameters from MCMC analysis with different combinations of datasets.}
\begin{tabular*}{\linewidth}{l@{\extracolsep{\fill}}cccc}
    \hline
    \hline
     & $\Lambda$CDM & Planck & Planck+SD & Planck+SD+D/H \\
    \hline
    $f_\gamma$ 
    & ... 
    & $ 0.66^{+0.26}_{-0.21}   $ 
    & $ 0.66^{+0.32}_{-0.17}    $ 
    & ... \\
    $\log_{10}(\Gamma_{Y}/\SI{}{\per\second})$ 
    & ... 
    & ...
    & $ > -5.72   $
    & $ > -6.15     $ \\
    $\log_{10}(T_{\mathrm{RH}}/ \mathrm{MeV})$ 
    & ... 
    & ...   
    & $ > -2.65   $
    & $ > -2.86     $ \\
    $\maxyr$ 
    & ... 
    & ... 
    & ...
    & $ < 0.0235    $\\
    $H_0 \,[\SI{}{\kilo\meter\per\second\per\mega\parsec}]$ 
    & $ 67.3^{+1.2}_{-1.2}   $ 
    & $ 66.2^{+2.8}_{-3.0}   $ 
    & $ 66.2^{+2.9}_{-3.0}    $
    & $ 67.4^{+1.7}_{-1.5}     $\\
    $\YHe$ 
    & $ 0.24683^{+0.00013}_{-0.00012}   $ 
    & $ 0.2508^{+0.0050}_{-0.0041}   $ 
    & $ 0.2508^{+0.0055}_{-0.0042}   $
    & $ 0.24696^{+0.00026}_{-0.00021}   $\\
    $10^5 \mathrm{D/H}$ 
    & $ 2.522^{+0.056}_{-0.056} $ 
    & $ 1.52^{+0.96}_{-0.93}   $ 
    & $ 1.58^{+0.92}_{-0.97}    $
    & $ 2.484^{+0.077}_{-0.087}     $\\
    $10^{-2}\omega_b$ 
    & $ 2.234^{+0.030}_{-0.029}  $ 
    & $ 2.225^{+0.040}_{-0.042}    $ 
    & $ 2.226^{+0.041}_{-0.043}   $
    & $ 2.232^{+0.030}_{-0.029}   $\\
    $\omega_{cdm}$ 
    & $ 0.1202^{+0.0027}_{-0.0027}  $ 
    & $ 0.1172^{+0.0065}_{-0.0067}    $ 
    & $  0.1172^{+0.0061}_{-0.0067}  $
    & $ 0.1208^{+0.0037}_{-0.0034}    $\\
    $100 \theta_s$ 
    & $ 1.04186^{+0.00058}_{-0.00058}   $ 
    & $ 1.0424^{+0.0013}_{-0.0011}   $ 
    & $ 1.0424^{+0.0013}_{-0.0011}   $
    & $ 1.04176^{+0.00066}_{-0.00075}   $\\
    $\ln 10^{10} A_s$ 
    & $ 3.045^{+0.033}_{-0.031}   $ 
    & $ 3.036^{+0.036}_{-0.037}   $ 
    & $ 3.037^{+0.036}_{-0.038}   $
    & $ 3.045^{+0.033}_{-0.031}  $\\
    $n_s$ 
    & $ 0.9642^{+0.0087}_{-0.0087}   $ 
    & $ 0.959^{+0.015}_{-0.016}   $ 
    & $ 0.959^{+0.015}_{-0.016}   $
    & $ 0.965^{+0.010}_{-0.0096}    $\\
    $\tau_{reio}$ 
    & $ 0.054^{+0.016}_{-0.015}   $ 
    & $ 0.053^{+0.015}_{-0.015}   $ 
    & $ 0.054^{+0.016}_{-0.015}    $
    & $ 0.054^{+0.016}_{-0.015}   $\\
    \hline
    \hline
\end{tabular*}
    \label{tab:posteriors}
\end{table*}

\begin{figure*}[t]
\centering
  \includegraphics[width=\linewidth]{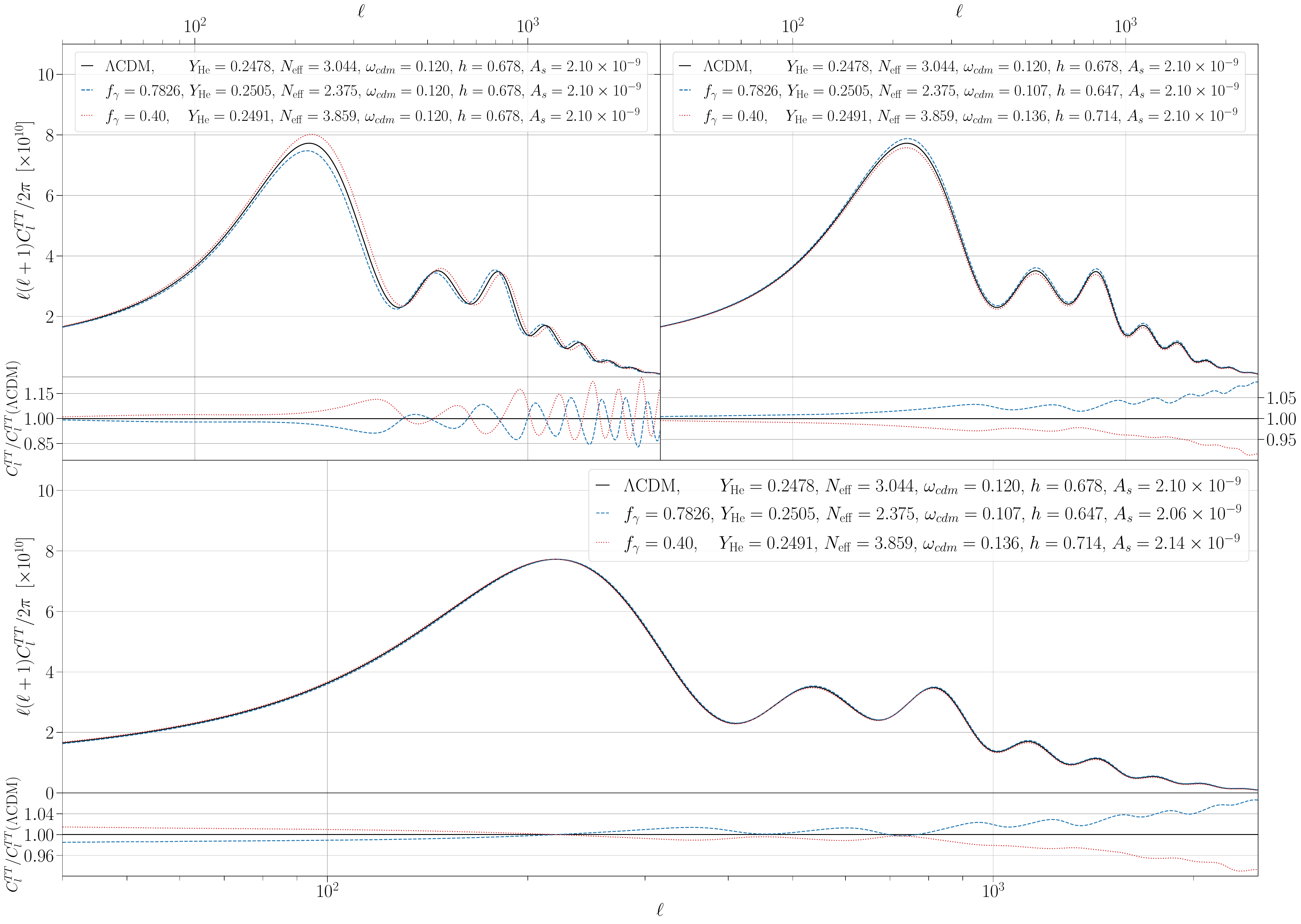}
  \caption{\footnotesize Power spectra for decays with $\maxyr = 0.2$, $T_{\mathrm{RH}} =\SI{3.16e-3}{MeV}$, and either $f_\gamma = 0.4$ (dotted red line) or $f_\gamma = 0.7826$ (dashed blue line). \textbf{Top left}: power spectra of two decay scenarios compared to spectrum of best-fit $\Lambda$CDM parameters (solid black line) with no other parameters being altered. \textbf{Top right}: $\omega_{cdm}$ and $h$ are simultaneously altered in order to fix $z_{eq}$ and $\theta_s$, respectively. \textbf{Bottom}: A change in $A_s$, in addition to $\omega_{cdm}$ and $h$, brings the spectra of these decays scenarios close to the best-fit $\Lambda$CDM spectrum. The remaining small-scale residuals are asymmetric; the $f_\gamma = 0.7826$ case is aided by the increase in $\YHe$ and therefore has a smaller residual than the $f_\gamma = 0.4$ case. }
  \label{fig:powerspectra_progression}
\end{figure*}

This process of fixing $z_{eq}$ and $\theta_s$ is illustrated in Fig.\ \ref{fig:powerspectra_progression}. The top left panel of Fig.\ \ref{fig:powerspectra_progression} depicts the power spectra for two decay scenarios described by $\maxyr = 0.2$, $T_{\mathrm{RH}} =\SI{3.16e-3}{MeV}$, and either $f_\gamma = 0.4$ (dotted red line) or $f_\gamma = 0.7826$ (dashed blue line). These values of $f_\gamma$ correspond to equivalent deviations in either direction about the sweet spot $f_\gamma = 0.5913$. Then, in the top right panel of Fig.\ \ref{fig:powerspectra_progression}, we alter $\omega_{cdm}$ in order to fix $z_{eq}$ and change $h \equiv H_0/(\SI{100}{\kilo\meter\per\second\per\mega\parsec})$ to compensate for the change in $\theta_s$. Even though $\theta_s$ is fixed, there is a lingering offset in peak locations. As mentioned in Sec.\ \ref{sec:Neff}, this phase shift is caused by the addition of extra free-streaming radiation \cite{bashinsky_signatures_2004,follin_first_2015,baumann_phases_2016}. Furthermore, the alteration of $\omega_{cdm}$ results in an amplification or suppression of all modes. Therefore, Planck data adjusts the amplitude $A_s$. The bottom panel of Fig.\ \ref{fig:powerspectra_progression} shows a continuation of the top right panel, with corrections to $\omega_{cdm}$ and $h$, but now with an additional change in $A_s$ such that the first peak height is the same as that of the best-fit $\Lambda$CDM spectrum. 

The lingering small-scale damping effects of $\Neff$ and $\YHe$ discussed in Sec.\ \ref{sec:Neff} are now seen in the bottom panel of Fig.\ \ref{fig:powerspectra_progression}. The case of $f_\gamma = 0.7826$ ($\Neff < 3.044$) has a lack of small scale damping compared to $\Lambda$CDM whereas the $f_\gamma = 0.4$ ($\Neff > 3.044$) case has too much damping on small scales. Changing the spectral index, $n_s$, can compensate for this non-standard Silk damping \cite{planck_collaboration_planck_2014}. Non-standard damping can also be compensated by changes in $\omega_b$ such that the number of free electrons increases or decreases and alters the mean free path of photons. However, changing $\omega_b$ has the unintended consequence of affecting the height ratio of odd and even peaks and is therefore a costly approach to accommodating a $Y$ decay scenario.

The corrections to $\omega_{cdm}$, $h$, $A_s$, $n_s$, and $\omega_b$, described above are apparent in the Planck 1D posteriors of Fig.\ \ref{fig:triangle}. However, note that there is an asymmetry in these corrections. The 1D posterior of $\omega_{cdm}$, for example, is not symmetric about the $\Lambda$CDM distribution. This asymmetry is due to the increase in $\YHe$ discussed in Sec.\ \ref{sec:YHe}. For $f_\gamma > 0.5913$, the decrease in $\Neff$ is partially compensated by an increase in $\YHe$. Therefore, decreasing $n_s$ or $\omega_b$ can compensate for values of $f_\gamma > 0.59$, making these scenarios more viable than the naive Planck bounds on $\Neff$ would imply.

On the other hand, if $f_\gamma < 0.5913$, then $\Neff > 3.044$, leading to too much damping on small scales. Planck will try to compensate by increasing $n_s$ or $\omega_b$. However, $\YHe$ is still larger compared to $\Lambda$CDM in this $f_\gamma < 0.5913$ regime. So, any increase in $n_s$ or $\omega_b$ is fighting against the combined damping effects of increasing both $\Neff$ and $\YHe$. Therefore, Planck tends to favor $f_\gamma > 0.59$ rather than $f_\gamma < 0.59$. 

This asymmetric effect can also be seen in the residuals of the bottom panel of Fig.\ \ref{fig:powerspectra_progression}. The case of $f_\gamma=0.7826$ has smaller residuals on small scales than that of the $f_\gamma=0.4$ scenario because the increased $\YHe$ is helping to compensate for the lack of damping caused by a decreased $\Neff$. The $f_\gamma - \maxyr$ plane in Fig.\ \ref{fig:triangle} further demonstrates the Planck preference for $f_\gamma > 0.5913$. When only Planck data is considered, $\maxyr$ is unconstrained as long as $f_\gamma$ is near the ``sweet spot'' value that maintains $\Neff\simeq 3.0$ after the $Y$-induced entropy injection. However, the Planck 1D posterior peaks at $f_\gamma = 0.632$ rather than $f_\gamma = 0.5913$ (gray dashed line), which would give $\Neff = 3.044$.

In theory, Planck would accept even larger $\maxyr$ at the sweet spot $f_\gamma$ until reaching a limit that corresponds to the Planck bounds on $\YHe$; \textit{Planck} 2018 high-$\ell$ TT,TE,EE+low-$\ell$ EE places a 2$\sigma$ upper bound of $\YHe = 0.283$ \cite{planck_collaboration_planck_2020}. Indeed, we obtain similar results when using a prior of $\maxyr = [0,1.2]$. However, exploring the $\maxyr$ corresponding to this $\YHe$ Planck upper bound is unnecessary since the addition of our D/H bounds greatly constrains $\YHe$.\footnote{ Additionally, the accuracy of the approximation in Eq.\ \eqref{eq:max_analitical} begins to diminish as $\maxyr\rightarrow 1$. See Appendix \ref{sec:CLASSmodifications}.} Since marginalizing over $f_\gamma$ leaves $\maxyr$ unconstrained for Planck and Planck+SD, we do not report an upper bound on $\maxyr$ for these likelihoods in Table \ref{tab:posteriors}. We also note that the steep dropoff seen at large $\YHe$ in the posteriors shown in Fig.\ \ref{fig:triangle_full} results from the MCMC sampling reaching the $\maxyr=0.8$ upper bound of our selected prior. 

The inclusion of SD constraints with Planck anisotropy data (Planck+SD) has the added power of restricting $T_{\mathrm{RH}}$ to hotter reheat temperatures, as expected. SDs constrain the reheat temperature to be $T_{\mathrm{RH}} > \SI{2.24e-3}{MeV} $ ($\Gamma_Y > \SI{1.91e-6}{\per\second}$) at $95\%$ C.L. Otherwise, SDs provide no additional constraining power on cosmological parameters. 

It is clear from Fig.\ \ref{fig:triangle} that the addition of a bound on D/H greatly constrains $Y$ decay scenarios compared to Planck+SD. First, the D/H constraint favors $f_\gamma \rightarrow 0$ to minimize changes in $\etaBBN$. This agrees with our assessment in Fig.\ \ref{fig:max_f_neff}. The combination of this preference for $f_\gamma \rightarrow 0$ with the Planck+SD 1D posterior for $f_\gamma$ that peaks around $f_\gamma \approx 0.632$ results in a relatively flat  $f_\gamma$ posterior for Planck+SD+D/H. 

The constraint on D/H disfavors large values of $\maxyr$. Non-zero values of $f_\gamma$ mean new photons will be injected by the $Y$ decay and therefore the photon density at BBN will be decreased accordingly. D/H prefers to minimize this effect by restricting $\maxyr$ to be less than 0.0235 (95\% C.L.). This limit on $\maxyr$ propagates to all other parameters; large values of $\YHe$ are no longer permitted and any corrections to the base six parameters that Planck favored are no longer needed because $\Neff$ is not deviating much from the standard value of $3.044$. Therefore, the full Planck+SD+D/H combination is in excellent agreement with $\Lambda$CDM values for the base six parameters \{$\omega_b$, $\omega_{cdm}$, $\theta_s$, $A_s$, $n_s$, $\tau_{reio}$\} and $H_0$ (see Table \ref{tab:posteriors}).

\begin{figure}[t]
\centering
  \includegraphics[width=\linewidth]{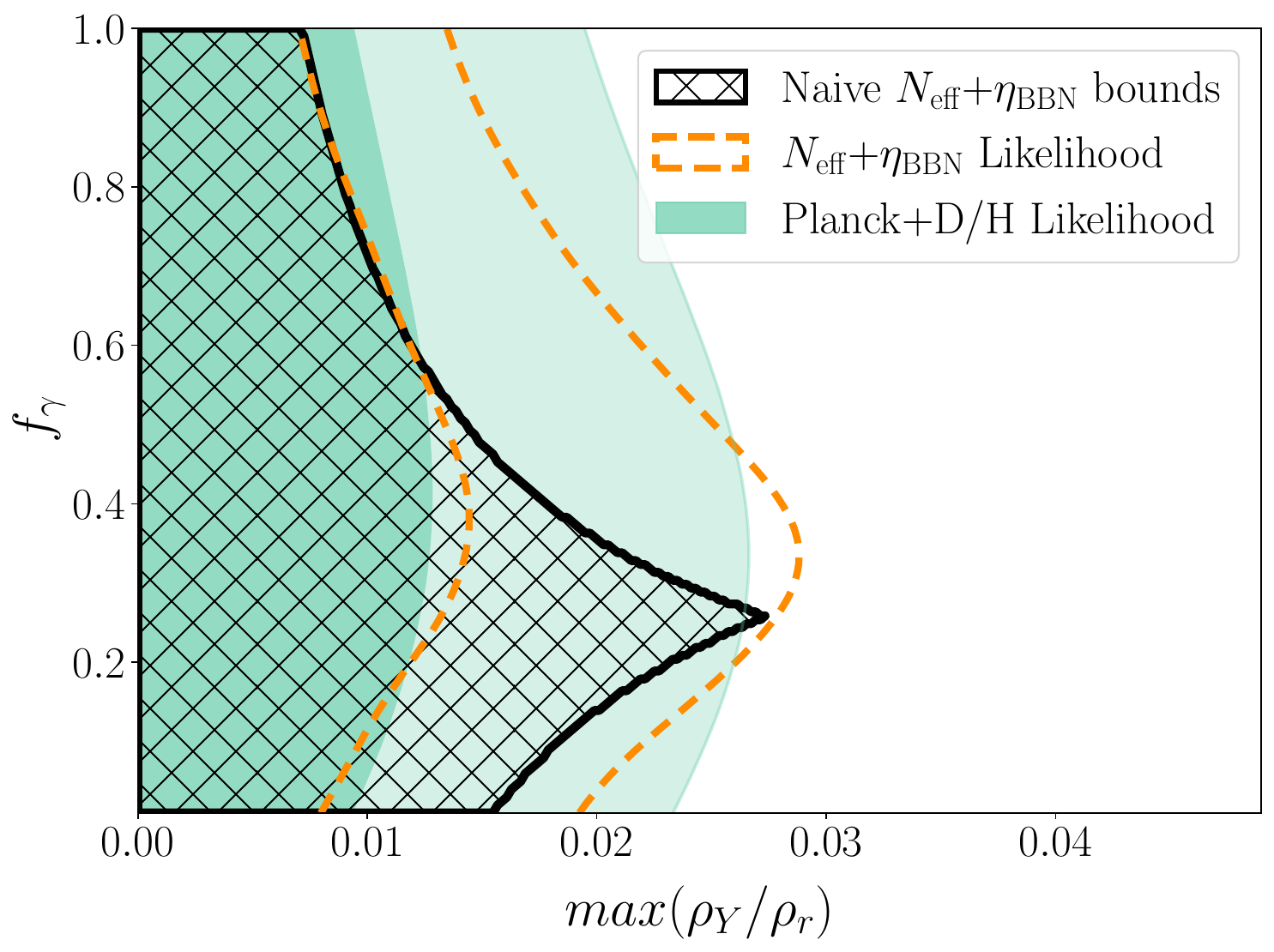}
  \caption{\footnotesize Comparison of the bounds on $f_\gamma$ and $\maxyr$ from different implementations of constraints. The hatched region is the overlap of the $\Neff$ and $\etaBBN$ contours in Fig.\ \ref{fig:max_f_neff} which held $\omega_b$ fixed. The dashed orange lines show the 68\% and 95\% contours from a Gaussian likelihood constructed with the 2018 \textit{Planck} bounds of $\Neff = 2.92^{+0.36}_{-0.37}$ and a Gaussian likelihood created with the bounds $\etaBBN = (6.119\pm0.100)\times10^{-10}$. Using the entire Planck+D/H likelihood (filled $68\%$ and $95\%$ contours) incorporates any uncertainty in cosmological parameters and the asymmetric preference Planck has for large $f_\gamma$.} 
  \label{fig:contour_comparison}
\end{figure}

The posterior of $\Gamma_Y$ for Planck+SD+D/H extends to smaller decay rates than that of Planck+SD. This is a symptom of the D/H bounds heavily constraining $\maxyr$. Small decay rates ($\Gamma_Y \lesssim 10^{-6}\,\SI{}{\per\second}$) were disfavored by Planck+SD because $\maxyr$ was free to vary up to a value of 0.8. However, by restricting $\maxyr$ to smaller values, the addition of D/H made smaller decay rates more probable. Therefore, the combination of SD+D/H constraints leads to a slightly broader posterior on $\Gamma_Y$ than that obtained from SDs alone. Otherwise, the addition of D/H has no influence on $\Gamma_Y$. While the combination of $\Gamma_Y$ and $\maxyr$ does influence how much the $Y$ particle increases the expansion rate during BBN, the D/H abundance is far more sensitive to changes in $\etaBBN$ than changes to $H$. 

While stringent, the addition of bounds on D/H does not
make constraints from Planck obsolete. Fig.\ \ref{fig:contour_comparison} shows a comparison of the 2D posterior bounds on $f_\gamma$ and $\maxyr$ for different implementations of constraints. The hatched region corresponds to the overlap between the $\Neff$ and $\etaBBN$ contours seen in Fig.\ \ref{fig:max_f_neff}. These limits were predicted using a fixed value of $\omega_b$, so this hatched region can be considered the most naive bounds on $f_\gamma$ and $\maxyr$. The dashed orange lines illustrate the 68\% and 95\% contours resulting from constraining $f_\gamma$ and $\maxyr$ with a Gaussian likelihood constructed with the 2018 \textit{Planck} bounds of $\Neff = 2.92^{+0.36}_{-0.37}$ and a Gaussian likelihood created with the bounds $\etaBBN = (6.119\pm0.100)\times10^{-10}$. These contours were also created for a fixed value of $\omega_b$. Here we see that marginalizing the bounds on $\Neff$ and $\etaBBN$ through an MCMC approach extends the allowed $\maxyr-f_\gamma$ parameter space compared to that of the naive bounds on $\Neff$ and $\etaBBN$. The filled contours in Fig.\ \ref{fig:contour_comparison} show the 2D posteriors for $\maxyr$ and $f_\gamma$ when constraining with a full Planck likelihood rather than just a Gaussian likelihood on $\Neff$, as well as Gaussian likelihood with $1\sigma$ bounds $(\text{D/H}) = (2.527\pm0.068 )\times10^{-5}$. The full Planck analysis not only accounts for uncertainties in cosmological parameters such as $\omega_b$ and $n_s$, but also includes the interplay between $\YHe$ and $\Neff$ discussed in Sec.\ \ref{sec:Neff}. The Planck preference for $f_\gamma \approx 0.632$ permits values of $f_\gamma$ that are otherwise ruled out by the bounds on just $\Neff$ and $\etaBBN$. Therefore, the effects of the $Y$ decay on $\Neff$ and $\YHe$ still manifest in the Planck+D/H posteriors even though $\maxyr$ is significantly restricted compared to the Planck posteriors.  

\begin{figure}[t]
\centering
  \includegraphics[width=\linewidth]{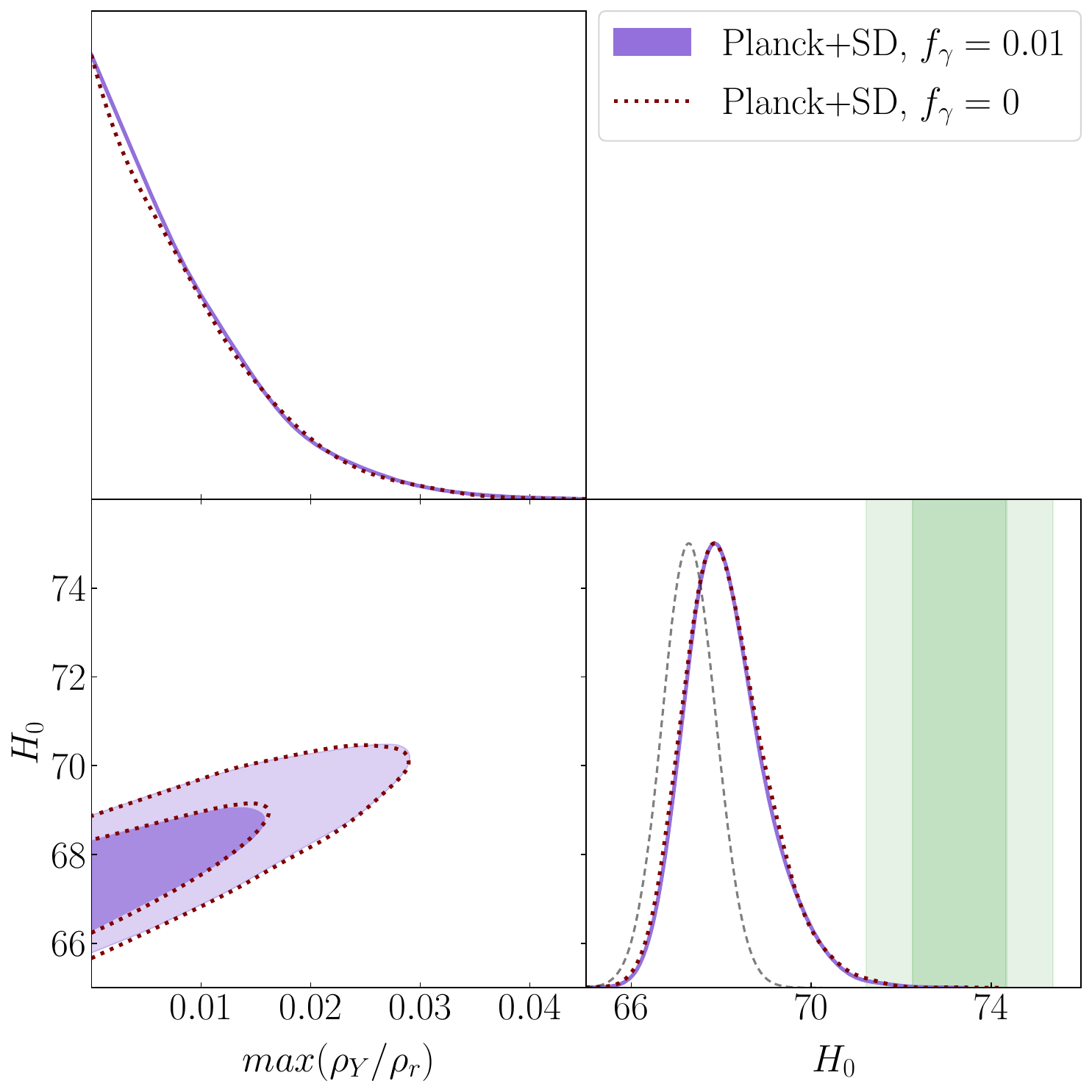}
  \caption{\footnotesize Comparison between $H_0$ posteriors of Planck+SD likelihood for $T_{\mathrm{RH}} = [9.5\times10^{-4}, 10^{-2}]$ MeV and either $f_\gamma = 0$ (dotted maroon outline) or $f_\gamma = 0.01$ (filled purple contour) with $68\%$ and $95\%$ C.L.\ contours. We include the 1D $H_0$ posterior for $\Lambda$CDM constrained by Planck (dashed gray line) and the vertical band shows the $1\sigma$ and $2\sigma$ bounds determined by S$H_0$ES ($H_0 = \SI{73.30 \pm 1.04}{\kilo\meter\per\second\per\mega\parsec}$). }
  \label{fig:fzero_H0}
\end{figure}

Finally, we note that even though this analysis considered the range of $f_\gamma = [0.01, 1]$, our results can be extended to smaller values of $f_\gamma$ for the limited range of $\Gamma_Y = [3.43\times10^{-7}, 3.80\times10^{-5}] \,\SI{}{\per\second}$. Fig.\ \ref{fig:fzero_f0p01} in Appendix \ref{sec:MCMCruns} demonstrates that the posteriors for all parameters are indeed identical when comparing $f_\gamma=0$ and $f_\gamma=0.01$ for $T_{\mathrm{RH}} = [9.5\times10^{-4}, 10^{-2}]$ MeV ( $\Gamma_Y = [3.43\times10^{-7}, 3.80\times10^{-5}] \,\SI{}{\per\second} $). We show a subset of these Planck+SD posteriors in Fig.\ \ref{fig:fzero_H0} and include the $1\sigma$ and $2\sigma$ bounds on $H_0$ (vertical bands) reported by the S$H_0$ES collaboration \cite{riess_comprehensive_2021}. When only considering Planck+SD, both the $f_\gamma=0$ and $f_\gamma=0.01$ cases favor values of $H_0$ larger than the best-fit $\Lambda$CDM value of $H_0 = \SI{67.3}{\kilo\meter\per\second\per\mega\parsec}$. While the addition of D/H bounds will restrict $\maxyr$ in the $f_\gamma=0.01$ case so that the $H_0$ posterior is in further disagreement with S$H_0$ES, $f_\gamma=0$ decays will not affect $\etaBBN$ and thus we do not expect the addition of D/H constraints to substantially change the Planck+SD constraints for $f_\gamma = 0$ seen in Fig.\ \ref{fig:fzero_H0}. For $f_\gamma=0$, the only additional constraints that D/H would place on $\maxyr$ would come from the increased expansion rate during BBN changing D/H. This change in $H$, however, has a small effect on D/H and can be reduced by decreasing $T_{\mathrm{RH}}$. Therefore, it is likely that the Planck+SD+D/H posteriors for $f_\gamma=0$ decays with colder reheat temperatures would not significantly differ from those for Planck+SD shown in Fig.\ \ref{fig:fzero_H0}. As mentioned in Sec.\ \ref{sec:distortion calculations}, investigating reheat temperatures $T_{\mathrm{RH}} \lesssim \SI{9e-4}{MeV} $ would require consideration of perturbations. It is not yet clear if the inclusion of $Y$ perturbations and corrections to the perturbations of ultra-relativistic species would relax the Planck+SD+D/H bounds on an \mbox{$f_\gamma=0$} decay. 

\section{Summary and Conclusions} \label{sec:conclusions}
In this work, we test the robustness of the common assumption of entropy conservation between BBN and recombination by considering the decay of a massive hidden sector $Y$ particle. We assume the hidden sector is sufficiently cold compared to the visible sector of Standard Model particles such that the particle is non-relativistic during BBN. These decay scenarios alter the effective number of relativistic species at recombination, $\Neff$, as well as the baryon-to-photon ratio and expansion rate at BBN. By employing observations of \textit{Planck} temperature and polarization anisotropies, CMB spectral distortions, and the primordial deuterium abundance, we determine constraints on the decay rate of the $Y$ particle ($\Gamma_Y$), its maximum contribution to the energy density of the Universe ($\maxyr$), and the fraction of the $Y$ particle’s energy density that is transferred to photons ($f_\gamma$).

If the $Y$ particle decays into a mixture of photons and other ultra-relativistic particles such as dark radiation, then there is a sweet spot photon fraction, $f_\gamma = 0.5913$, that will keep $\Neff$ fixed at the fiducial 3.044. However, the injection of photons from the $Y$ particle leads to an increase in $\etaBBN$ due to the requirement that $T_0 = \SI{2.7255}{\kelvin}$ (see Appendix \ref{sec:CLASSmodifications}). This increase in $\etaBBN$ increases $\YHe$ which results in \textit{Planck} observations of the CMB temperature and polarization anisotropies preferring a larger photon fraction, with the 1D posterior for $f_\gamma$ peaking at 0.632. For $f_\gamma$ between 0.555 and 0.707, CMB anisotropies permit significant entropy injections in which the energy density of the $Y$ particle equals at least half of the radiation density prior to its decay. If $f_\gamma$ deviates from the sweet spot value, then cosmological parameters like $\omega_{cdm}$, $H_0$, $A_s$, $\omega_b$, and $n_s$ must be altered from their standard $\Lambda$CDM values in order to match the observed \textit{Planck} temperature power spectrum. 

Upper limits on $\mu$ and $y$ spectral distortions of the CMB restrict the decay rate of the $Y$ particle to $\Gamma_Y=\SI{1.91e-6}{\per\second}$ (95\% C.L.) such that maximum lifetime of the $Y$ particle is $\SI{5.25e5}{\sec}$. The Primordial Inflation Explorer, which proposes to measure the CMB blackbody spectrum with three orders of magnitude better sensitivity than the results of COBE/FIRAS \cite{kogut_primordial_2011}, would enhance bounds on $\mu$ and $y$ distortions and therefore place even more restrictive bounds on the lifetime of the particle. 

The deuterium abundance, D/H, proves to have the most constraining power on $Y$-induced entropy injection. We apply bounds of $\mathrm{D/H} = (2.527 \pm 0.068)\times10^{-5}$, which we translate from bounds on the baryon-to-photon ratio reported by \textcite{cooke_one_2018}. This uncertainty in D/H includes measurement uncertainty and uncertainties in nuclear reaction rates.\footnote{After our analysis was complete, \textcite{yeh_probing_2022} released a new computation of the deuterium abundance that employs updated cross-section measurements from the LUNA Collaboration \citep{mossa_baryon_2020}. Assuming a fixed value of $\eta$ and $N_{\mathrm{eff}} = 3.044$, \textcite{yeh_probing_2022} report values for $\sigma_{\mathrm{DH}}$ due to uncertainties in three reaction rates. These uncertainties combine to give a theoretical uncertainty in D/H of $7.5\times10^{-7}$. When combined with the measurement uncertainty of $3.0\times10^{-7}$, this result gives $\sigma_{\mathrm{DH}} = 8.1\times10^{-7}$, which is less stringent than the D/H bound used in this work. } This bound on deuterium limits the energy density of the $Y$ particle to be no more than 2.35\% (95\% C.L.) the total energy density of radiation. Using the fit described in Fig.\ \ref{fig:gap_maxYR}, this upper bound corresponds to $g < 1.05$, which translates to a 5\% increase in the comoving energy density of radiation. Limits on D/H will likely tighten as more absorption systems are observed and as laboratory measurements of nuclear reaction rates improve. 

Photodisintegration of light nuclei by the injection of photons can generate even tighter constraints on entropy injection scenarios. \textcite{balazs_cosmological_2022} recently performed an analysis in which they explored the constraints that could be placed on axion-like particles with masses greater than 2.2 MeV decaying into photons between BBN and recombination by employing CMB anisotropy and spectral distortion data as well as detailed BBN constraints via $\etaBBN$ and photodisintegration of nuclei. The addition of photodisintegration restricts the maximum energy density of the axion-like particle, $\rho_a$, to be $\rho_{a}/\rho_{tot} < 10^{-3}$. A decaying particle with a mass below 3.2 MeV evades these photodisintegration limits; the stringent BBN bounds on $\maxyr$ derived here only consider the effects of a $Y$ decay on $\etaBBN$ and the expansion rate during BBN. We find $\maxyr < 0.0235$ (95\% C.L.), demonstrating that a small level of entropy injection could be possible in the regime of a sub-$3.2$ MeV decay for which photodisintegration bounds are irrelevant.

Even though bounds on deuterium proved to be the most restrictive on a general entropy injection scenario, we demonstrated that there exist subtleties in the dependence of the decay on degeneracies between $\Neff$ and $\YHe$ that require a full CMB analysis (Figure \ref{fig:contour_comparison}). Furthermore, we investigated the effects that entropy injection has on standard cosmological parameters, including $H_0$. In doing so, we find that any injection that includes photons is heavily restricted by observations of the CMB and D/H such that $H_0 = 67.4^{+1.7}_{-1.5}\,\SI{}{\kilo\meter\per\second\per\mega\parsec}$ ($95\%$ C.L.).

We marginalized results over the range of $0.01 \leq f_\gamma \leq 1$ and therefore we cannot speak with absolute confidence regarding an $f_\gamma = 0$ injection scenario. If $f_\gamma = 0$, then spectral distortions would not restrict the $Y$ particle lifetime (see Fig.\ \ref{fig:mu_constraints}), allowing for reheat temperatures at which the scales accessible to \textit{Planck} enter the horizon. It would therefore be necessary to consider perturbation dynamics for the $Y$ particle and how its decay products affect the evolution of other perturbations. However, we are able to extend our results to $f_\gamma=0$ for a limited range of reheat temperatures; when considering $T_{\mathrm{RH}} = [9.5\times10^{-4}, 10^{-2}]$ MeV ( $\Gamma_Y = [3.43\times10^{-7}, 3.80\times10^{-5}] \,\SI{}{\per\second} $), we can neglect any influences the $Y$ particle has on perturbations. For this limited range of reheat temperatures, we show that the posteriors for $f_\gamma=0$ and $f_\gamma=0.01$ are identical when constraining with \textit{Planck} anisotropies and spectral distortions (Figure \ref{fig:fzero_f0p01}). While the addition of bounds on D/H would restrict $\maxyr$ in the $f_\gamma =0.01$ case, decays with $f_\gamma = 0$ would leave $\etaBBN$ unaltered and only minimally affect the deuterium abundance by increasing the Hubble rate during BBN. Therefore, scenarios with the $Y$ particle decaying fully into dark radiation are less restricted than the $f_\gamma \neq 0$ cases studied in this work. Such decays into dark radiation increase $H_0$ and begin to mitigate the Hubble tension. When considering only \textit{Planck} anisotropies and spectral distortions, $f_\gamma = 0$ decays in this limited range of decay rates result in $H_0 = 68.1^{+1.9}_{-1.8}\,\SI{}{\kilo\meter\per\second\per\mega\parsec} $ ($95\%$ C.L.), which only slightly reduces the Hubble tension. As was demonstrated by the progression of early dark energy models \cite{karwal_early_2016,poulin_early_2019,smith_oscillating_2020}, however, the inclusion of perturbations can lead to significant changes in results. For long-lived $Y$ scenarios, altering perturbations could potentially relax the bounds on the injection of dark radiation and alleviate the Hubble tension even further. We reserve an in-depth investigation of the $f_\gamma = 0$ case for a future work. 

We have demonstrated that the injection of new radiation between BBN and recombination, be it photons or other ultra-relativistic species such as dark radiation, is highly constrained: a massive particle whose decay products include photons can make up at most $2.35\%$ of the energy density of the Universe. This stringent limit implies that cosmological 
parameters, including the Hubble constant, cannot be adjusted by introducing changes in the photon abundance between BBN and recombination. Our results uphold the standard cosmological picture and the assumption that entropy is conserved.

\section*{Acknowledgments} \label{sec:Acknowledgments}

This analysis employed the Longleaf Computing Cluster owned by the University of North Carolina at Chapel Hill. A.C.S.\ and A.L.E.\ are supported in part by NSF CAREER Grant No.\ PHY-1752752. A.C.S.\ also acknowledges support from the NC Space Grant Consortium and the Bahnson Fund at UNC Chapel Hill. T.L.S.\ is supported by NSF Grant No.~2009377, NASA Grant No.~80NSSC18K0728, and the Research Corporation.

\appendix

\section{INITIAL CONDITIONS FOR CLASS}\label{sec:CLASSmodifications}
Here we describe how to recast our set of descriptive parameters for the $Y$ decay model into a set of initial conditions. We start by making the approximation that $H(a) \approx H_i (a/a_i)^{-2}$ during radiation domination (RD). This assumption, along with taking $\rho_Y = \rho_{Y,i}$ when $a=a_i$, allows us to analytically solve Eq.\ \eqref{eq:rho_Y}:
\begin{align}
    \rho_Y(a) = \rho_{Y,i} \left(\frac{a_i}{a}\right)^{3} e^{\frac{1}{2} \widetilde{\Gamma}_Y\left[1-(a/a_i)^2\right]} \,. \label{eq:rho_Y_analytical}
\end{align}
Additionally, we define the reheat scale factor by $\Gamma_Y \equiv H_i(a_i/a_{\mathrm{RH}})^{2}$: $a_{\mathrm{RH}}/a_i = \widetilde{\Gamma}_Y^{-1/2}$, where $\widetilde{\Gamma}_Y \equiv \Gamma_Y/H_i$. It follows from Eq.\ \eqref{eq:rho_Y_analytical} that the maximum value for $\rho_Y/\rho_r$ occurs at $a_{\mathrm{RH}}/a_i$. Therefore, an analytical expression for $\maxyr$ can be found by evaluating Eq.\ \eqref{eq:rho_Y_analytical} at $a_{\mathrm{RH}}$ and taking $\rho_r(a_{\mathrm{RH}}) = \rho_{r,i} (a_i/a_{\mathrm{RH}})^{4}$:
\begin{align}
    max\left(\frac{\rho_Y}{\rho_r}\right) = e^{\frac{1}{2}(\tilde{\Gamma}_Y-1)} \frac{\rho_{Y,i}}{\rho_{r,i}} \,\,\tilde{\Gamma}_Y^{-1/2}  \,. \label{eq:max_analytical_appendix}
\end{align}
Eq.\ \eqref{eq:max_analytical_appendix} defines the $\maxyr$ decay parameter used throughout this work. This analytical expression for $\maxyr$ deviates from the exact numerical maximum of $\rho_{Y}/\rho_r$ as $\maxyr$ approaches unity. Comparing Eq.\ \eqref{eq:max_analytical_appendix} to the numerically calculated $\maxyr$ for various decay scenarios, we find that Eq.\ \eqref{eq:max_analytical_appendix} is accurate within $1\%$ for $\maxyr < 0.079$ and within $10\%$ for $\maxyr < 0.945$. Seeing that a value of $\maxyr > 1$ translates to a period $Y$ domination, it is consistent that our radiation domination approximation would begin fail as $\maxyr$ approaches 1. 

In order to determine a value for $\rho_{Y,i}$ from Eq.\ \eqref{eq:max_analytical_appendix}, we need to find the initial condition $\rho_{r,i} = \rho_{\gamma,i}+ \rho_{ur,i}+\rho_{ncdm,i}$. For decay scenarios with $f_\gamma \neq 0$, we must decrease the photon energy density pre-decay such that the photon injection from the $Y$ decay results in a present day temperature of $T_0 = \SI{2.7255}{\kelvin}$. Since we only consider reheat temperatures after BBN ($T_{\mathrm{RH}} \lesssim \SI{0.01}{MeV}$), we must decrease $\rho_\gamma$ at the time of neutrino decoupling and therefore we must also rescale the initial energy density of ultra-relativistic species, $\rho_{ur,i}$. To do so, we parametrize the initial energy densities by $\Omega'_{x,0}$ such that $\rho_{x,i} = \Omega'_{x,0} \rho_{\text{crit},0} \, a_i^{-4}$ for species $x$. In Eq.\ \eqref{eq:gap}, we define the ratio of comoving radiation energy density before and after decay as 
\begin{align}
    g \equiv \frac{\rho_{r,f}a_f^4}{\rho_{r,i}a_i^4} \,,
\end{align}
where the $i$ and $f$ subscripts denote before and after decay, respectively. This $g$ parameter is entirely dependent on $\maxyr$ defined by Eq.\ \eqref{eq:max_analytical_appendix}. Let us denote the comoving energy density of species $x$ as $\bar{\rho}_{x,j} = \rho_{x,j} a_j^4$. Then the change in comoving radiation density is
\begin{align}
    \Delta \bar{\rho}_r = \bar{\rho}_{r,f} - \bar{\rho}_{r,i} = \left(g-1\right) \bar{\rho}_{r,i} \,.
\end{align}
From this, we can specify the change in individual species with the photon fraction as
\begin{align}
    \Delta \bar{\rho}_{\gamma} &= f_\gamma \left(g-1\right) \bar{\rho}_{r,i} \,,\\
    \Delta \bar{\rho}_{ur} &= \left(1-f_\gamma\right) \left(g-1\right) \bar{\rho}_{r,i}\,.
\end{align}
Now we can determine the $g$ factor for each species that receives energy injection from the decay:
\begin{align}
        g_\gamma &= \frac{\bar{\rho}_{\gamma,f}}{\bar{\rho}_{\gamma,i}} = \frac{\bar{\rho}_{\gamma,i} + \Delta\bar{\rho}_\gamma}{\bar{\rho}_{\gamma,i}} \nonumber \\
        &= 1 + \frac{\Delta\bar{\rho}_\gamma}{\bar{\rho}_{\gamma,i}} = 1 + f_\gamma\left(g-1\right) \frac{\rho_{r,i}}{\rho_{\gamma,i}} \,,\\
        g_{ur} &= \frac{\bar{\rho}_{ur,f}}{\bar{\rho}_{ur,i}} = \frac{\bar{\rho}_{ur,i} + \Delta\bar{\rho}_{ur}}{\bar{\rho}_{ur,i}} \nonumber \\
        &= 1 + \frac{\Delta\bar{\rho}_{ur}}{\bar{\rho}_{ur,i}} = 1 + \left(1-f_\gamma\right) \left(g-1\right) \frac{\rho_{r,i} }{\rho_{ur,i}} \,.
\end{align}
We assume the minimal convention in which neutrinos are composed of two massless species ($ur$) and one massive with $m_{ncdm} = 0.06$ eV. The massive neutrino, or non-cold dark matter (ncdm), will be relativistic at early times and therefore contribute to the initial radiation energy density. If we denote $N_{ur}$ as the contribution to $\Neff$ from the ultra-relativistic species and $N_{ncdm}$ as the contribution from the massive neutrino, then we enforce that $N_{ur} + N_{ncdm} = 3.044$ at the time of BBN. With these definitions, we can write
\begin{align}
    \frac{\rho_{r,i}}{\rho_{\gamma,i}} & = \frac{\rho_{ncdm,i} + \rho_{ur,i} + \rho_{\gamma,i}}{\rho_{\gamma,i}} = \frac{\rho_{ncdm,i}}{\rho_{\gamma,i}} + \frac{\rho_{ur,i}}{\rho_{\gamma,i}} + 1 \nonumber \\
    &= \frac{7}{8} N_{ncdm} \left(\frac{4}{11} \right)^{4/3}\!\!\!+ \frac{7}{8} N_{ur} \left(\frac{4}{11} \right)^{4/3}\!\!\!+1  \nonumber \\
    &= \frac{7}{8} (3.044) \left(\frac{4}{11} \right)^{4/3}\!\!\!+ 1 \,\,,
\end{align}
and
\begin{align}
    \frac{\rho_{r,i}}{\rho_{ur,i}}  &= \frac{\rho_{ncdm,i} + \rho_{ur,i} + \rho_{\gamma,i}}{\rho_{ur,i}} = \frac{\rho_{ncdm,i}}{\rho_{ur,i}} + 1 +  \frac{\rho_{\gamma,i}}{\rho_{ur,i}} \nonumber \\
    &= \frac{N_{ncdm}}{N_{ur}} + 1 + \frac{8}{7}\frac{1}{N_{ur}} \left( \frac{11}{4}\right)^{4/3} \nonumber\\
    &= \frac{3.044 - N_{ur}}{N_{ur}} + 1 + \frac{8}{7}\frac{1}{N_{ur}} \left( \frac{11}{4}\right)^{4/3} \nonumber\\
    &= \frac{1}{N_{ur}} \left[ 3.044 + \frac{8}{7} \left( \frac{11}{4}\right)^{4/3}  \right] \,\,. 
\end{align}
It follows that the ratio of comoving energy density for each species is 
\begin{align}
    g_\gamma &= 1 + f_\gamma\left(g-1\right) \left[1 + \frac{7}{8} (3.044) \left(\frac{4}{11} \right)^{4/3} \right]\,\,, \label{eq:gap_gamma_appendix} \\
    g_{ur} &= 1 + \frac{(1-f_\gamma)\left(g-1\right)}{N_{ur}} \left[ 3.044 + \frac{8}{7} \left( \frac{11}{4}\right)^{4/3} \right] \,\,.\label{eq:gap_ur_appendix} 
\end{align}
If $\Omega_{\gamma,0}$ is the photon energy density corresponding to $T_0 = \SI{2.7255}{\kelvin}$, then the initial values are found by $\rho_{x,i} = \Omega'_{x,0} \rho_{\text{crit},0} \, a_i^{-4}$ and 
\begin{align}
\Omega'_{\gamma,0} &= \frac{\Omega_{\gamma,0}}{g_\gamma} = \frac{\Omega_{\gamma,0}}{1 + f_\gamma\left(g-1\right) \left[1 + \frac{7}{8} (3.044) \left(\frac{4}{11} \right)^{4/3} \right]} , \label{eq:Omega_g_prime} \\
\Omega'_{ur,0} &= \frac{\Omega_{ur,0}}{g_{ur}} = \frac{7}{8}  N_{ur} \left(\frac{4}{11} \right)^{4/3} \Omega'_{\gamma,0} \,\,. \label{eq:Omega_ur_prime}
\end{align}
In Eq.\ \eqref{eq:Omega_ur_prime}, $\Omega_{ur,0} = (7/8)  N_{ur}^{post} \left(4/11 \right)^{4/3} \Omega_{\gamma,0}$, where $N_{ur}^{post}$ is the effective number of ultra-relativistic species after decay (see Appendix \ref{sec:postdecay}). Since $g$ is purely a function of $\maxyr$ (Figure \ref{fig:gap_maxYR}), we can determine the initial conditions $\rho_{\gamma,i}$ and $\rho_{ur,i}$ directly from our decay parameters $f_\gamma$ and $\maxyr$. In the case of $f_\gamma = 0$ (no decays into photons), we do not need to decrease $\rho_\gamma$ at early times. Indeed, plugging $f_\gamma = 0$ into Eq.\ \eqref{eq:Omega_g_prime} leads to $\Omega'_{\gamma,0} = \Omega_{\gamma,0}$ and the initial condition is simply $\rho_{\gamma,i} = \Omega_{\gamma,0}\rho_{crit}a_i^{-4}$. The contribution to $\Neff$ from the ultra-relativistic neutrino species before decay is determined by $N_{ur} = 3.044 - N_{ncdm}$, where $N_{ncdm}$ is the contribution that the massive neutrino makes to $\Neff$ prior to recombination.

Massive neutrino (ncdm) calculations performed by CLASS derive the number density, energy density, and pressure of massive neutrinos based on the input temperature $T_{ncdm,0}/T_{\gamma,0}$. By default, CLASS assumes this temperature to be $T_{ncdm,0} = 0.71611 \,T_{\gamma,0}$ in order to obtain a mass-to-density ratio of $m/\omega_{ncdm} = 93.14$ eV. However, this default value inherently assumes that both $T_{ncdm}$ and $T_{\gamma}$ scale as $T \propto a^{-1}$ after electron-positron annihilation. While this scaling is indeed true for $T_{ncdm}$, any $Y$ decay scenario that injects photons will result in $T_{\gamma}$ not consistently scaling as $a^{-1}$. The photon temperature at some scale factor, $a_i$, which occurs between electron-positron annihilation and energy injection from the $Y$ decay, will depend on the initial photon energy density determined by Eq.\ \eqref{eq:Omega_g_prime}. Specifically,
\begin{align}
   T_{\gamma,i} a_{i} &= \left(\frac{\Omega'_{\gamma,0}}{\Omega_{\gamma,0}} \right)^{1/4}\!\!\! T_{\gamma,0} a_0,
\end{align}
where $a_0$ is the present-day scale factor. Assuming that $T_{ncdm,i} = 0.71611 \,T_{\gamma,i}$ after electron-positron annihilation and before the $Y$ particle alters the evolution of $T_\gamma$, 
\begin{align}
    T_{ncdm,0} &= 0.71611 \left(\frac{\Omega'_{\gamma,0}}{\Omega_{\gamma,0}} \right)^{1/4}\!\!\! T_{\gamma,0} \,\,. 
\end{align}
If there is no photon injection from the decay, then $\Omega'_{\gamma,0} = \Omega_{\gamma,0}$ and we recover the default assumption of CLASS. We emphasize that this change in the energy density of massive neutrinos is a result of the $Y$ decay altering the scaling of $T_\gamma$. As discussed in Sec.\ \ref{sec:model}, we assume that the $Y$ decay does not produce any new active neutrinos.

The contribution that the massive neutrino makes to $\Neff$, which we denote as $N_{ncdm}$, is
\begin{equation}
    N_{ncdm} = \left( \frac{0.71611}{(4/11)^{1/3}} \right)^{4} = 1.0132 \,\,.
\end{equation}
This value is used to determine the number of ultra-relativistic species pre-decay, $N_{ur}$, by enforcing $3.044 = N_{ur} + N_{ncdm}$ at BBN.

\section{POST-DECAY \texorpdfstring{$\Neff$}{TEXT} }\label{sec:postdecay}
We define the post-decay $\Neff$ to be the number of relativistic species immediately after entropy injection from the $Y$ decay has completed. For the reheat temperatures that we consider in this work, this means that massive neutrinos are still relativistic and therefore contribute to $\Neff$ immediately after decay. As discussed in Sec.\ \ref{sec:CLASSmodifications}, the contribution that the massive neutrino makes to $\Neff$  before the decay is $N_{ncdm} = (11/4)^{4/3} (0.71611)^4 =1.0132$. While the $Y$ particle does not inject any new massive neutrinos, the ratio of $\rho_{ncdm}/\rho_\gamma$ will change due to the decay creating new photons. Therefore, the contribution that the massive neutrino makes to $\Neff$ will change post-decay. This post-decay number of massive neutrinos is 
\begin{align}
    N_{ncdm}^{post} = \frac{N_{ncdm}}{g_\gamma} = \left(\frac{11}{4} \right)^{4/3} \frac{(0.71611)^4}{g_\gamma} \,\,,
\end{align}
where $g_\gamma$ is defined by Eq.\ \eqref{eq:gap_gamma_appendix}. Additionally, the effective number of ultra-relativistic species will change due to the $Y$ decay. If $N_{ur}$ is the effective number of relativistic species \textit{before} decay, then we denote the contribution of the ultra-relativistic species to the post-decay $\Neff$ as $N_{ur}^{post}$. It follows that
\begin{align}
    \Omega_{ur,0} & = \Omega'_{ur} g_{ur}  \nonumber\\
    \frac{7}{8} N_{ur}^{post} \left(\frac{4}{11} \right)^{4/3} \Omega_{\gamma,0} & = \frac{7}{8} N_{ur} \left(\frac{4}{11} \right)^{4/3} \Omega'_{\gamma,0} g_{ur}  \nonumber\\
    N_{ur}^{post} &= \frac{\Omega'_{\gamma,0}}{\Omega_{\gamma,0}} g_{ur} N_{ur} \nonumber\\
    N_{ur}^{post} &= \frac{ g_{ur} }{g_\gamma} N_{ur} \,\,,
\end{align}
where $g_{ur}$ is defined by Eq.\ \eqref{eq:gap_ur_appendix}. Therefore, the total post-decay $\Neff$ is 
\begin{align}
    \Neff^{post} &= N_{ncdm}^{post} + N_{ur}^{post} = \frac{1}{g_\gamma} \left[\left(\frac{11}{4} \right)^{4/3} (0.71611)^4  +  g_{ur} N_{ur}  \right] \nonumber \\
    &= \frac{1}{g_\gamma} \left[ 3.044 + N_{ur}(g_{ur} - 1) \right] \,\,,
\end{align}
where $N_{ur} = 2.0308$. Since both $g_{\gamma}$ and $g_{ur}$ are ultimately functions of $\maxyr$ and $f_\gamma$, we are able to calculate the post-decay $\Neff$ directly from our decay parameters.

\onecolumngrid
\clearpage
\section{ADDITIONAL MCMC RESULTS} \label{sec:MCMCruns}
\begin{figure*}[b]
\centering
  \includegraphics[width=\linewidth]{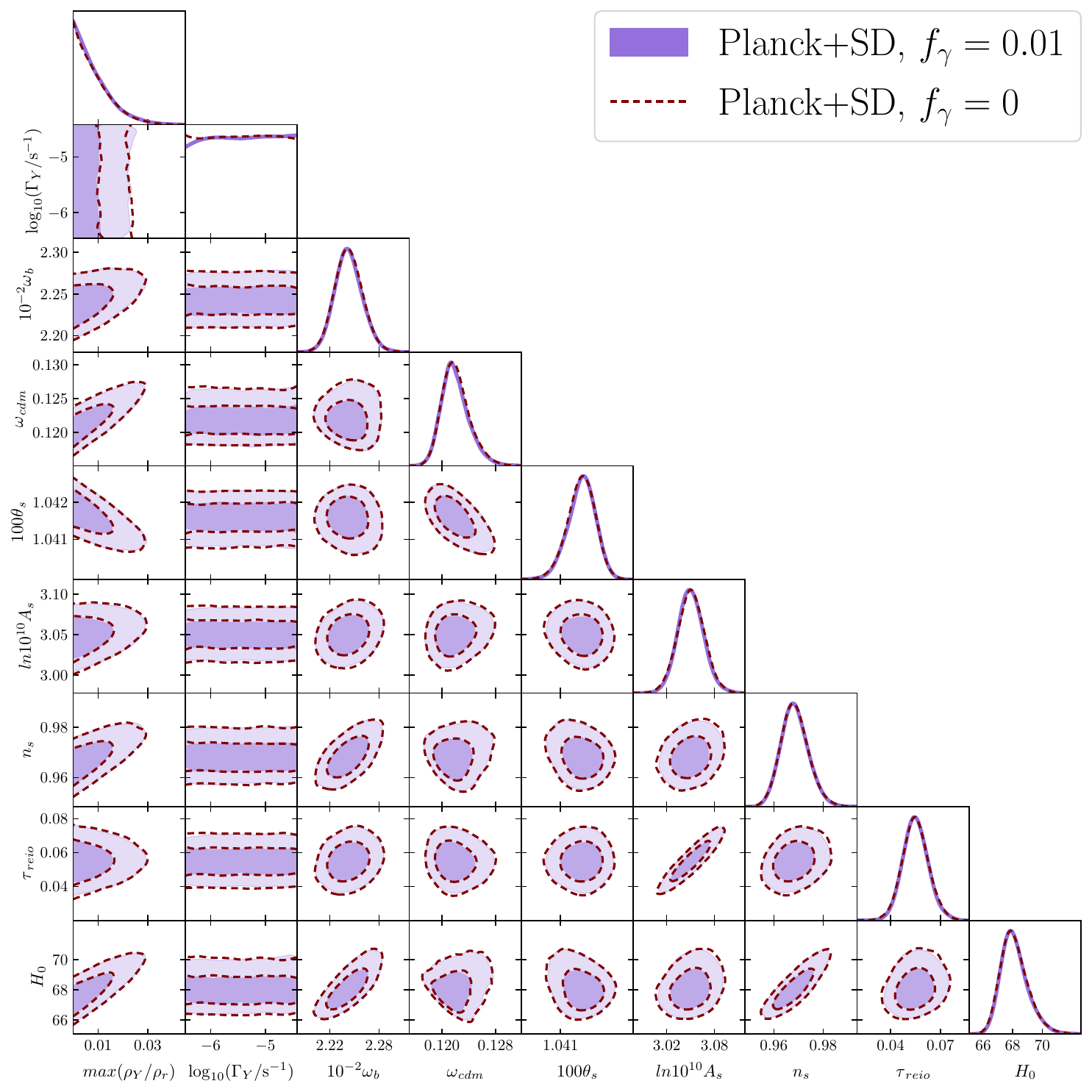}
  \caption{\footnotesize Comparison between posteriors of Planck+SD likelihood for $T_{\mathrm{RH}} = [9.5\times10^{-4}, 10^{-2}]$ MeV and either $f_\gamma = 0$ (dashed outline) or $f_\gamma = 0.01$ (filled) with $68\%$ and $95\%$ C.L.\ contours. The posteriors of $f_\gamma = 0$ vs. those of $f_\gamma = 0.01$ are identical in all parameters except for a slight difference in $\Gamma_Y$. The $f_\gamma = 0$ case is completely unconstrained by SDs and therefore equally favors long and short particle lifetimes whereas the $f_\gamma = 0.01$ case disfavors long lifetimes due to the constraints placed by SDs. }
  \label{fig:fzero_f0p01}
\end{figure*}

\begin{figure*}[b]
\centering
  \includegraphics[width=\linewidth]{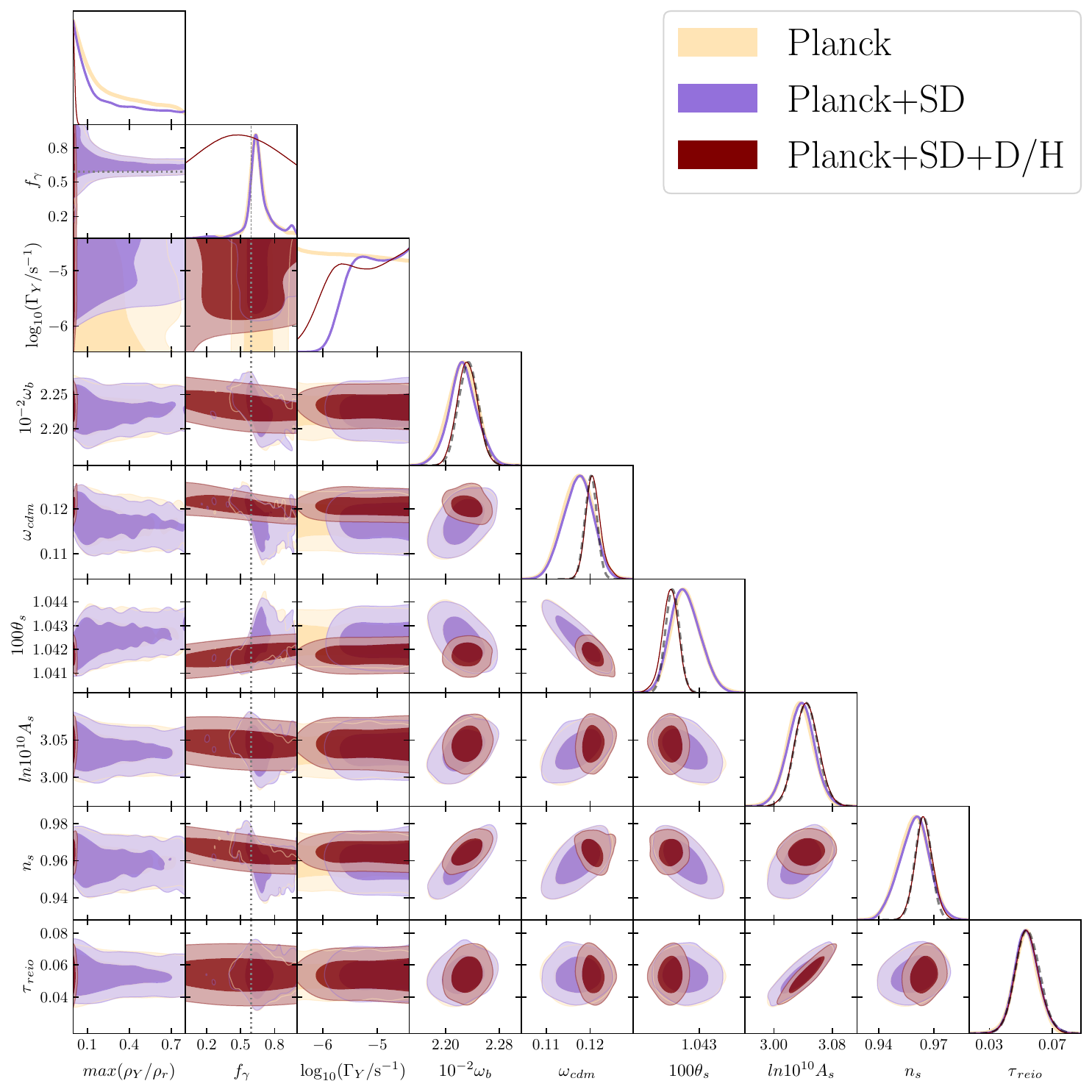}
  \caption{\footnotesize 1D and 2D posterior distributions of decay and full set of cosmological parameters for different combinations of \textit{Planck} high-$\ell$ TT,TE,EE, low-$\ell$ TT, and low-$\ell$ EE  (\textbf{Planck}) data, CMB spectral distortions (\textbf{SD}) bounds, and constraints on the deuterium abundance (\textbf{D/H}). We include the 1D posteriors for $\Lambda$CDM constrained by Planck (dashed black line). The dotted gray line traces $f_\gamma = 0.5913$, which maintains $\Neff = 3.044$ at recombination.}
  \label{fig:triangle_full}
\end{figure*}

\clearpage

\twocolumngrid
\bibliography{Was_Entropy_Conserved_Between_BBN_and_Recombination}{}

\end{document}